\documentclass[acmtocl]{acmtrans2m}

\usepackage{amsfonts}
\usepackage{amsmath}
\usepackage{amssymb}
\usepackage{amsbsy}
\usepackage{proof}
\usepackage{diagrams}

\newdimen\proofrulebreadth \proofrulebreadth=.05em
\newdimen\proofdotseparation \proofdotseparation=1.25ex
\newdimen\proofrulebaseline \proofrulebaseline=2ex
\newcount\proofdotnumber \proofdotnumber=3
\let\then\relax
\def\hfi{\hskip0pt plus.0001fil}
\mathchardef\squigto="3A3B
\newif\ifinsideprooftree\insideprooftreefalse
\newif\ifonleftofproofrule\onleftofproofrulefalse
\newif\ifproofdots\proofdotsfalse
\newif\ifdoubleproof\doubleprooffalse
\let\wereinproofbit\relax
\newdimen\shortenproofleft
\newdimen\shortenproofright
\newdimen\proofbelowshift
\newbox\proofabove
\newbox\proofbelow
\newbox\proofrulename
\def\shiftproofbelow{\let\next\relax\afterassignment\setshiftproofbelow\dimen0 }
\def\shiftproofbelowneg{\def\next{\multiply\dimen0 by-1 }%
\afterassignment\setshiftproofbelow\dimen0 }
\def\setshiftproofbelow{\next\proofbelowshift=\dimen0 }
\def\setproofrulebreadth{\proofrulebreadth}

\def\prooftree{
\ifnum  \lastpenalty=1
\then   \unpenalty
\else   \onleftofproofrulefalse
\fi
\ifonleftofproofrule
\else   \ifinsideprooftree
        \then   \hskip.5em plus1fil
        \fi
\fi
\bgroup
\setbox\proofbelow=\hbox{}\setbox\proofrulename=\hbox{}%
\let\justifies\proofover\let\leadsto\proofoverdots\let\Justifies\proofoverdbl
\let\using\proofusing\let\[\prooftree
\ifinsideprooftree\let\]\endprooftree\fi
\proofdotsfalse\doubleprooffalse
\let\thickness\setproofrulebreadth
\let\shiftright\shiftproofbelow \let\shift\shiftproofbelow
\let\shiftleft\shiftproofbelowneg
\let\ifwasinsideprooftree\ifinsideprooftree
\insideprooftreetrue
\setbox\proofabove=\hbox\bgroup$\displaystyle 
\let\wereinproofbit\prooftree
\shortenproofleft=0pt \shortenproofright=0pt \proofbelowshift=0pt
\onleftofproofruletrue\penalty1
}

\def\eproofbit{
\ifx    \wereinproofbit\prooftree
\then   \ifcase \lastpenalty
        \then   \shortenproofright=0pt  
        \or     \unpenalty\hfil         
        \or     \unpenalty\unskip       
        \else   \shortenproofright=0pt  
        \fi
\fi
\global\dimen0=\shortenproofleft
\global\dimen1=\shortenproofright
\global\dimen2=\proofrulebreadth
\global\dimen3=\proofbelowshift
\global\dimen4=\proofdotseparation
\global\count255=\proofdotnumber
$\egroup  
\shortenproofleft=\dimen0
\shortenproofright=\dimen1
\proofrulebreadth=\dimen2
\proofbelowshift=\dimen3
\proofdotseparation=\dimen4
\proofdotnumber=\count255
}

\def\proofover{
\eproofbit 
\setbox\proofbelow=\hbox\bgroup 
\let\wereinproofbit\proofover
$\displaystyle
}%
\def\proofoverdbl{
\eproofbit 
\doubleprooftrue
\setbox\proofbelow=\hbox\bgroup 
\let\wereinproofbit\proofoverdbl
$\displaystyle
}%
\def\proofoverdots{
\eproofbit 
\proofdotstrue
\setbox\proofbelow=\hbox\bgroup 
\let\wereinproofbit\proofoverdots
$\displaystyle
}%
\def\proofusing{
\eproofbit 
\setbox\proofrulename=\hbox\bgroup 
\let\wereinproofbit\proofusing
\kern0.3em$
}

\def\endprooftree{
\eproofbit 
  \dimen5 =0pt
\dimen0=\wd\proofabove \advance\dimen0-\shortenproofleft
\advance\dimen0-\shortenproofright
\dimen1=.5\dimen0 \advance\dimen1-.5\wd\proofbelow
\dimen4=\dimen1
\advance\dimen1\proofbelowshift \advance\dimen4-\proofbelowshift
\ifdim  \dimen1<0pt
\then   \advance\shortenproofleft\dimen1
        \advance\dimen0-\dimen1
        \dimen1=0pt
        \ifdim  \shortenproofleft<0pt
        \then   \setbox\proofabove=\hbox{%
                        \kern-\shortenproofleft\unhbox\proofabove}%
                \shortenproofleft=0pt
        \fi
\fi
\ifdim  \dimen4<0pt
\then   \advance\shortenproofright\dimen4
        \advance\dimen0-\dimen4
        \dimen4=0pt
\fi
\ifdim  \shortenproofright<\wd\proofrulename
\then   \shortenproofright=\wd\proofrulename
\fi
\dimen2=\shortenproofleft \advance\dimen2 by\dimen1
\dimen3=\shortenproofright\advance\dimen3 by\dimen4
\ifproofdots
\then
        \dimen6=\shortenproofleft \advance\dimen6 .5\dimen0
        \setbox1=\vbox to\proofdotseparation{\vss\hbox{$\cdot$}\vss}%
        \setbox0=\hbox{%
                \advance\dimen6-.5\wd1
                \kern\dimen6
                $\vcenter to\proofdotnumber\proofdotseparation
                        {\leaders\box1\vfill}$%
                \unhbox\proofrulename}%
\else   \dimen6=\fontdimen22\the\textfont2 
        \dimen7=\dimen6
        \advance\dimen6by.5\proofrulebreadth
        \advance\dimen7by-.5\proofrulebreadth
        \setbox0=\hbox{%
                \kern\shortenproofleft
                \ifdoubleproof
                \then   \hbox to\dimen0{%
                        $\mathsurround0pt\mathord=\mkern-6mu%
                        \cleaders\hbox{$\mkern-2mu=\mkern-2mu$}\hfill
                        \mkern-6mu\mathord=$}%
                \else   \vrule height\dimen6 depth-\dimen7 width\dimen0
                \fi
                \unhbox\proofrulename}%
        \ht0=\dimen6 \dp0=-\dimen7
\fi
\let\doll\relax
\ifwasinsideprooftree
\then   \let\VBOX\vbox
\else   \ifmmode\else$\let\doll=$\fi
        \let\VBOX\vcenter
\fi
\VBOX   {\baselineskip\proofrulebaseline \lineskip.2ex
        \expandafter\lineskiplimit\ifproofdots0ex\else-0.6ex\fi
        \hbox   spread\dimen5   {\hfi\unhbox\proofabove\hfi}%
        \hbox{\box0}%
        \hbox   {\kern\dimen2 \box\proofbelow}}\doll%
\global\dimen2=\dimen2
\global\dimen3=\dimen3
\egroup 
\ifonleftofproofrule
\then   \shortenproofleft=\dimen2
\fi
\shortenproofright=\dimen3
\onleftofproofrulefalse
\ifinsideprooftree
\then   \hskip.5em plus 1fil \penalty2
\fi
}

\newtheorem{definition}{Definition}[section]
\newtheorem{lemma}[definition]{Lemma}
\newtheorem{proposition}[definition]{Proposition}
\newtheorem{remark}[definition]{Remark}

\newtheorem{theorem}[definition]{Theorem}
\newtheorem{corollary}[definition]{Corollary}

\newcommand{\Set}[1]{\{#1\}}
\newcommand{\Nat}{\mathbb{N}} 
\newcommand{\Bool}{\mathbb{B}} 
\newcommand{\tuple}[1]{\langle #1 \rangle} 
\newcommand{\TrueVal}{\mbox{\sf true}}
\newcommand{\FalseVal}{\mbox{\sf false}}

\newcommand{\Const}[1]{{\sf #1}}
\newcommand{\Succ}{\Const{succ}}

\newcommand{\union}{\;\cup\;}
\newcommand{\intersection}{\;\cap\;}

\newcommand{\Iff}{\Leftrightarrow}

\newcommand{\arrow}{\rightarrow}

\newcommand{\LangZero}{{\cal L}_0} 
\newcommand{\LangOne}{{\cal L}_1} 

\newcommand{\Sem}[1]{[\hspace{-0.6mm}[ #1 ]\hspace{-0.6mm}]}

\newcommand{\State}             { { \mathbb{S} } }

\newcommand{\Id}                {{\tt Id}}

\newcommand{\der}{\vdash}

\newcommand{\BoldText}[1]{\mbox{\bf #1}}

\newcommand{\CatC}{{\cal C}}

\newcommand{\CatSet}{\BoldText{Set}}

\newcommand{\PRA}{\mbox{\bf PRA}}
\newcommand{\IPC}{\mbox{\bf IPC}} 
\newcommand{\CPC}{\mbox{\bf CPC}} 
\newcommand{\HA}{\mbox{\bf HA}}
\newcommand{\MA}{\mbox{\bf MA}}
\newcommand{\PA}{\mbox{\bf PA}}
\newcommand{\EM}{\mbox{\bf EM$_1$}}

\newcommand{\Then}{\rightarrow}
\newcommand{\Or}{\vee}
\newcommand{\Not}{\neg}
\newcommand{\fv}[1]{\mbox{\rm FV}(#1)}
\newcommand{\States}{\mathbb{S}}
\newcommand{\join}{\sqcup}
\newcommand{\sleq}{\sqsubseteq}

\newcommand{\Cat}{{\cal C}}
\newcommand{\Compatible}[2]{#1 \uparrow #2}
\newcommand{\Prefix}[1]{\mbox{\it Prefix}\,(#1)}

\newcommand{\forces}[3]{#1 \Vdash #2:#3}

\newcommand{\ext}[1]{\mbox{\it ext}(#1)}

\newcommand{\SetC}{\ensuremath{\bf Set}}
\newcommand{\SMonad}{{\cal S}}
\newcommand{\SSunit}{\eta^{\cal S}}
\newcommand{\Sunit}{\eta}
\newcommand{\SSext}[1]{#1^{*^{\cal S}}}
\newcommand{\Sext}[1]{#1^{*}}

\newcommand{\RMonad}{{\cal R}}
\newcommand{\Runit}{\eta^{\cal R}}
\newcommand{\Rext}[1]{#1^{*^{\cal R}}}

\newcommand{\EMonad}{{\cal E}}
\newcommand{\Eunit}{\eta^{\cal E}}
\newcommand{\Eext}[1]{#1^{*^{\cal E}}}

\newcommand{\natTrans}{\stackrel{.~}{\rightarrow}}
\newcommand{\sym}[1]{{\sf #1}}
\newcommand{\chiP}{\chi_{\sym{P}}}

\newcommand{\phiP}{\varphi_{\sym{P}}}
\newcommand{\phiQ}{\varphi_{\sym{Q}}}

\newcommand{\Pred}[1]{\sym{#1}}
\newcommand{\Fun}[1]{\sym{#1}}

\newcommand{\DefEq}{:=}
\newcommand{\monoid}{\,\otimes}

\newcommand{\Smonoid}{\monoid^{\SMonad}}

\newcommand{\NOT}{\stackrel{\cdot}{\neg}}
\newcommand{\AAND}{\stackrel{\cdot}{\wedge}}
\newcommand{\OOR}{\stackrel{\cdot}{\vee}}
\newcommand{\IMPL}{\stackrel{\cdot}{\rightarrow}}
\newcommand{\STAR}{\stackrel{\cdot}{\star}}

\newcommand{\Lbot}{\lambda\_\,.\bot}

\newcommand{\BibTeX}{{\rm B\kern-.05em{\sc i\kern-.025em b}\kern-.08em
    T\kern-.1667em\lower.7ex\hbox{E}\kern-.125emX}}

\title{Interactive Realizers and Monads}
\author{STEFANO BERARDI\\Universit\`a di Torino\\
UGO DE'LIGUORO\\Universit\`a di Torino}

\begin{abstract}
We propose a realizability interpretation of a system for quantiÞer free arithmetic which is equivalent to the fragment of classical arithmetic without {\em nested} quantiÞers, called here $\EM$-arithmetic. We interpret classical proofs as interactive learning strategies, namely as processes going through several stages of knowledge and learning by interacting with the ``environment'' and with each other. We give a categorical presentation of the interpretation through the construction of two suitable monads.
\end{abstract}


{}






\begin{document}

\begin{bottomstuff}
Author's address: S. Berardi and U. de'Liguoro, Dipartimento di Informatica, Universit\`a di Torino, c.so Svizzera 185, 10149 Torino, Italy. Phone +39 011 6706711 - fax +39 011 751603.
E-mail address: {\tt $\{$stefano,deligu$\}$@di.unito.it}
\end{bottomstuff}
\maketitle


\section{Introduction}
\label{sec:introduction}

Since the discovery, by G\"odel, Gentzen, Kreisel and others, of
interpretations of the classical into the intuitionistic logic, it
has been held that classical proofs should have a constructive
significance, which logicians and computer scientists may envisage
as their computational content. This soon appeared, and still
remains, one of the more promising fields of application of logic
to mathematics and computer science, with a positive contribution
by logic rather than the finding of negative results only.

The impressive amount of work by several authors carried out so
far has produced a rich deal of suggestive ideas and results
supporting them; however, in our opinion, a convincing account of
the nature of such a content is still missing, either because of
its indirect description through translations and interpretations,
or because the proposed approaches are based on almost magic
properties of formal systems (think of G\"odel's dialectica
interpretation and of its usage together minimal logic, or
Friedman's A-translation), or because of the too fine
grained analysis, often at the level of logical connectives and
quantifiers on which they rely. The discomfort with such
descriptions of the constructive content of classical proofs
becomes more evident when looking at the ``algorithms'' one
obtains by means of the known techniques, which are, worse than
inefficient, impossible to follow and likely far away from the
ideas on which the proofs were based.

We wish to go a further step into a recently emerged approach,
which in our view originates from the game theoretic account of
logic and arithmetic, and especially Coquand's semantics of
evidence in \cite{Coq95}. We look for an explicit
description of the constructive meaning of proofs, possibly
relating the obtained algorithms to the proof ideas they should
come from. The basic intuition is that of understanding proofs as
strategies, by which a finite agent learns how to ensure the truth
of the proof conclusion. We emphasize here the point of {\em
learning in the limit} as the distinctive feature of the approach
we are developing. Learning is an interactive process in which
(general) hypothesis are raised and tested against (particular)
facts, possibly realizing that some wrong guesses have been made
and reacting in some clever way. Also learning is an intrinsically
unbounded process, in the sense that its goal is never definitely
attained: or at least, the learning agent could not be able to
decide to have been eventually successful.

A rough account of this follows the lines of Tarski semantics of
formulas and of Kleene realizability interpretation of
quantifiers, but for the crucial case of the existential
quantification. As a basic assumption let us take for granted that
quantifier free sentences (hence without any occurrence of
variables, neither bound nor free) can be checked against truth by
direct computation. A strategy for $\forall x\,A(x)$ is a uniform
method telling how to learn about $A(c)$ for any possible choice
of the (individual denoted by) the constant $c$. The learning of
the truth of $\exists x\,A(x)$, instead, involves the guessing of
a $c$ such that $A(c)$ can be learned to be true, without being
$c$ computable in general. By this reason a guess cannot be
regarded as a definite choice, rather as a tentative one: if and
when an evidence should occur against its correctness, the learner
has to be instructed on how to backtrack her guesses and try some new guess.

Consider for concreteness the excluded middle law, {\bf EM}, that
we write in the suggestive form: $\exists x\,A(x) \vee \forall y\,
\neg A(y)$. Rephrasing in the learning scenario Coquand's dialogic
interpretation of this law, the learning of {\bf EM} begins by
assuming $\forall y\, \neg A(y)$, which is by the above the
strategy of learning $\neg A(c)$ for any arbitrary choice of $c$.
If an evidence occurs that $A(c')$ should be true for some
specific $c'$, then the learner changes her mind and backtracks to
the point where the previous assumption was made; she takes now
$A(c')$ as her new guess suspending at the same time her assent to
all the consequences that have been drown from the incompatible
assumption that $\neg A(c')$.

This process involves a memory, recording of all the previous
guesses and of the logical dependencies among them, and it could
be very complex depending on the logical complexity of $A$. If it
is a quantifier free closed formula, however, $A$ is decidable by
our basic assumption, so that the discovery that $A(c')$ actually
holds for some $c'$ is definite, and all the consequences drown
from its negation can be simply forgotten: this is what is called
1-backtracking in \cite{BerCoqHay}. In the case of a more complex
formula $A$ the learner might realize to have been wrong in
discarding $\forall y\,\neg A(y)$, and even to have been wrong in
her belief to have been wrong: in other words she might be in
place to resume her own rejected guesses, but also to resume her
negative attitude toward some of them, and so on.

A learning process of this guise is not guaranteed to terminate in
principle, and it would be contradictory with the undecidability
of the Halting problem to ask that, even if the learning process comes out from a
sound proof, it should reach a perfect knowledge within a finite
number of steps and effectively. It is at this point that we
resort to Gold's idea of learning in the limit (see
\cite{Gold65,Gold67}). A sound learning strategy (a winning
strategy in the game theoretic jargon) should guarantee that the
learner can be wrong in her guessing only finitely many times, so
that she will eventually hold her final guesses without any
further change of her mind because of the discovery of some
counterexample. The trick is that, while the process of generating
and discarding the guesses is effective, one accepts that, except in particular cases, 
it is undecidable whether the definite guess has been
reached or not.

Learning in the limit in the sense of Gold is not sufficient for
interpreting proofs of the whole classical arithmetic. It
corresponds to the 1-backtracking fragment which we call here
$\PRA+\EM$, and which is essentially Hayashi's Limit Computable
Mathematics \cite{Hay06}. We have studied in
\cite{Berardi-deLiguoro:IC09} the concept of limit in the general
case of unbounded backtracking, thought always of well-founded
depth; but a precise description of the interpretation of formal
proofs into learning strategies is a quite challenging task,
especially because we do not want to alter the form of the
conclusion, nor the proof itself by translating and forcing them
into some normal form, because classically equivalent formulas and
proofs might embody different constructive ideas.

Being aware of the difficulty, in this paper we address the issue
in the limited case of the quantifier free fragment of Heyting
arithmetic, $\HA$, known as primitive recursive arithmetic,
$\PRA$. As it is explained in the next section, we add to such a
theory the existential quantification only to the extent of
expressing $\EM$, which is {\bf EM} when $\exists x\, A(x)$ is
$\Sigma^0_1$, that is when $A$ is decidable. We do this by adding
Skolem functions to the language $\LangZero$ of $\PRA$, in a way
which is reminiscent of Hilbert's $\varepsilon$-terms, so that one
does not need to consider nested quantifiers in the technical
development. We then rephrase $\EM$ by means of new axioms
implying that, for each primitive recursive predicate $\Pred{P}$
of arity $k+1$ there is a $k$-ary function symbol $\phiP$ such
that
\[\Pred{P}(\vec{x},y) \rightarrow  \Pred{P}(\vec{x},\phiP(\vec{x})).\]
The effect of $\phiP$ is to choose a $y$ making
$\Pred{P}(\vec{x},y)$ true if such a $y$ exists depending on the
parameters $\vec{x}$; $\phiP(\vec{x})=0$ (or any other default
value) otherwise. But the function denoted by $\phiP$ does not
need to be computable; we view its values as guesses for $y$
instead, and accept the idea that the individual denoted by a term
including the symbols $\phiP$ might change while the learner's
finite knowledge of the standard  model grows.

We represent a state of knowledge by a finite set of atomic closed
formulas $\Pred{P}(\vec{m},n)$ which have been found to be true by
the learner in the standard recursion theoretic model of $\PRA$,
and such that they uniquely define a finite part of the relative Skolem functions
$\phiP$. Therefore by simply taking finite supersets of some given
state of knowledge as its extensions, we see that there are in
general many incompatible ways of enlarging the learner knowledge,
which correspond to the possible choices for defining the guessing
functions $\phiP$. To account for the dependency of the meaning of
terms and formulas on the state of knowledge we interpret the
individuals and the statements of the theory into functions from states to natural
numbers and booleans respectively. Because of this a formula which
was deemed false in some state $s$ might become true in some state
$s'\supseteq s$; but note that also the opposite is possible.

The way out from such a seemingly chaotic situation is the redefinition of the
concept of individual as a dynamic object. An {\em individual} $\alpha$, also
called a {\em strongly convergent function}, is a mapping from the
set of states $\State$ to some set of values $X$, such that given
any countable sequence of states $s_0 \subseteq s_1 \subseteq s_2
\subseteq \cdots$ which is weakly increasing w.r.t. the extension
relation, $\alpha(s_i)$ is eventually constant. An individual is a dynamic
or perhaps an epistemic concept, since it clearly evolves along
the history of a learning process depending on the actual
experiments made by the learner. In the case of a formula $A$ a more concrete description of
the learning strategy is a searching procedure that, given some
initial state $s_0$, produces consistent extensions $s_1, s_2,
\ldots$ of it, such that the meaning of $A$, which is a
function in $\State\arrow\Bool$ (where $\Bool$ is the set of truth
values) eventually becomes true.

We prove that under a natural lifting of the standard
interpretation of terms and formulas from the language
$\LangZero$, into functions from $\State$ to $\Nat$ and from $\State$ to
$\Bool$, terms and formulas always denote dynamic individuals, provided that the
variables occurring in them are also interpreted by individuals in
the new sense. 
We eventually obtain the desired result of having the whole language
$\LangOne$ of the theory $\PRA+\EM$ uniformly interpreted into the
same kind of mathematical objects.

Describing the model in terms of type theory, individuals inhabit
the types of the shape $\SMonad X = \State\arrow X$; the meaning
of $\SMonad$ is that of a {\em strong monad} in the sense of
\cite{Mog91}, which we call the {\em state monad}, that can be
seen as a type constructor. If we take for simplicity the category
of sets as our base interpretation category, then $\SMonad$ is an
endofunctor of $\CatSet$ for which there exist an (injective)
inclusion $\SSunit_X:X\arrow\SMonad X$ and an extension map
$\SSext{\_}:(X\arrow\SMonad Y) \arrow (\SMonad X\arrow\SMonad Y)$,
satisfying a suitable universality condition. $\SMonad X$
does contain more than individuals, but we characterize the functions
of the shape $\SSext{f}$ as those treating the state as a
consistent approximation of the definition of truth, by using only
the value $\alpha(s)$ in the evaluation of $\SSext{f}(\alpha,s)$.
Since $\alpha$ is the denotation of a term $t$ or of a formula
$A$, this amounts to evaluate all the parts of $t$ and $A$ into
the same $s$, which is a sort of global state: by this we say that a
function $\SSext{f}$ has a {\em global state}, and
call it {\em global} for short. Global functions have the pleasant
property of sending individuals to individuals, and to be
determined by their behaviour over the image of $\SSunit_X(X)$ into
$\SMonad X$. The
desired result of having terms and formulas denoting individuals
is obtained by interpreting function, predicate and even
connective symbols into global functions.

The relation with 1-backtracking is apparent from the fact
that an individual might change its value only finitely many times
along some given sequence of states of knowledge; this is further
clarified by the fact that for each tuple of numbers $\vec{m}$ the
value of $\phiP(\vec{m})$ is definitely set to $n$ by  adding the
information $\Pred{P}(\vec{m},n)$ to the current state; but we
have to allow for a finite number of changes, and not just one,
because the arguments of $\phiP$ might well be variables (hence
dynamic individuals) and because of the possible nesting of the
$\phiP$ symbols, by which the tuple of arguments $\vec{m} =
\vec{\alpha}(s)$ (for a vector $\vec{\alpha}$ of individuals and
some newly reached state $s$) in $\phiP(\vec{m})$ might also change.

The interpretation of the language $\LangOne$ into individuals is
just a first ingredient of the model; indeed the natural question
arises of what can we say about the meaning of $A$ when
$\PRA+\EM\der A$. A simple minded answer is that it is an individual
always converging to true. But this is not the case: indeed it is not difficult
to find an $A$ and  an $s\in\State$ such that $\PRA+\EM\der A$ and $A$ is false
in $s$: then the sequence $\sigma(i) = s$ for all $i\in\Nat$ is weakly 
increasing, and $A$ is definitely false along $\sigma$. The problem cannot be overcome
by restricting to strictly increasing sequences, since the state $s$ can be
extended by adding information which is irrelevant to $A$ (say $\Pred{P}(\vec{m},n)$
for a predicate $\Pred{P}$ not occurring in $A$). A subtler property holds instead: if
$\PRA+\EM\der A$ then for any $s$ we can effectively find an extension $s'\supseteq s$ such that
$A$ is true in $s'$, namely the subset of states in which $A$ is true is cofinal in $\State$ w.r.t.
the extension ordering. 

The main result of the present paper is a semi-constructive proof of this claim\footnote{A constructive proof can be given, however, and it can be used to produce effective bounds to the time complexity of the extracted algorithms. See \cite{Aschieri11}.}. 
Suppose that $\PRA+\EM\der A$.
Given an arbitrary initial state $s_0$ we construct a weakly increasing sequence $\sigma = s_0, s_1,\ldots$ such that
$A$ is eventually true along $\sigma$. The construction is a learning process which is however not
blind search: it
is the proof of $A$ that embodies the searching strategy,
which is at the same time continuous to depend only on a
finite amount of information, coherent to produce sensible
answers, and strongly convergent to ensure that the desired goal
of making $A$ true will be eventually reached.

An {\em interactive realizer} is a function $r\in \SMonad (\State)
= \State\arrow\State$ which satisfies the above requirements to
interpret a proof. By profiting of the improved exposition of the
model in \cite{AschieriB09} w.r.t. its very first presentation in
\cite{Berardi05} and \cite{Berdel08}, we see a realizer $r$ as computing a state $r(s)$
which is compatible with $s$ and includes only what is needed to
proceed toward the validation of a formula $A$. Suppose that the
free variables of $A(x)$ are just $x$ (an inessential restriction), and let
$\alpha\in\SMonad\Nat$ be the individual interpreting $x$ given by
some environment mapping $\xi:{\it Var} \arrow \SMonad \Nat$; then
we say that $r$ {\em forces} $\alpha$ into $A$, and write
$\forces{r}{\alpha}{A}$, if for any $s\in\State$, if $r(s)$ is the
empty set (the trivial state $\bot$) then $A(\alpha(s))$ is true
in $s$. We call the subset of such states the prefix points of
$r$, written $\Prefix{r}$, since the definition we have adopted
immediately implies that $r(s) = \bot$ if and only if $r(s)
\subseteq s$. The existence of a search by means of $r$ out of an
arbitrary starting state $s_0$ implies that $\Prefix{r}$ is
cofinal in $\State$, which is the way we understand the definite
catch of the values of $\alpha$ into the set $\Set{n\in\Nat\mid A(n)\; \mbox{is true}}$.

There is a subtle difficulty with this concept of forcing, which
otherwise could be confused with the homonym Kripke's relation
between possible worlds and formulas. We have observed above that
even if the knowledge grows, the best that we can hope is that a
formula which is a theorem of $\PRA+\EM$ will become eventually
true along certain (weakly) increasing sequence of states, that are
comparable to the branches of a tree-form Kripke model. This is
clearly weaker than Kripkes's forcing, which is monotonic along
the paths of a model formed by the accessibility relation; since a
statement is valid in a Kripke model if it is forced at the root
of the tree of possible worlds, it is forced by every possible
world in the model. Also terms take a fixed denotation in Kripke
models as soon as they come to ``existence'' in a possible world,
so that they have been also called ``rigid designators'' in
\cite{Kr:NN}. We observe that this exactly mirrors the
intuitionistic view of the validity of existential assertions,
whose construction is, according to the Brouwer-Heyting-Kolmogorow
interpretation, a pair of an individual (in the standard sense)
and of a constructive proof that it satisfies a given property,
where the individual part, as well the proof, must be known at
once.

In the case of our model, on the contrary, the denotations of
terms and formulas change with the state; but the state is the
only way for a realizer to force $\alpha$ into the goal 
$A$: it is like pointing at some target which is moved
by a side effect of the shots themselves, that can be eventually hit only through
an interactive process of trials and errors. The interaction,
therefore, depends on factors that are independent of the proof,
like the given $\alpha$ and the starting state of knowledge: this
is our account of why the interpretation of a non constructive
proof of $A$ is an unpredictable (morally non-deterministic)
process, whose actual behavior depends on the interaction with the
``environment'', namely any other proof using $A$ as a lemma and
interacting with the same state.

Coming back to the interpretation of proofs we now sketch how a
realizer can be constructed out of them. In our
model the forcing relation relative to some statement $A$ is  a
binary relation between realizers and dynamic individuals, so that
it is included into $\SMonad(\State)\times \SMonad(\Nat)$: we
consider $\RMonad X =
\SMonad(\Nat)\times\SMonad(\State)$ for historical reasons and
because of the isomorphism with the object part of the side effect monad
 $\EMonad X =
\State \arrow (\Nat\times\State)$. Indeed it turns out that
$\RMonad$ is itself a monad, whose extension operation $\Rext{\_}$ is defined by means of a
binary operation $\Smonoid$ over $\SMonad(\State)$, allowing  
to combine two realizers $r$ and $r'$
into a new one $r\Smonoid r'$. 
We call the so obtained realizer the {\em merge} of $r$ and $r'$, which
is nothing else than the resulting interaction between $r$ and $r'$, where
each of them is engaged in satisfying its own goal, namely the
premises of an inference rule in the proof.
This operation can be constructed in
several ways, but it is axiomatically definable as the lifting to
$\SMonad (\State)$ of a binary operation $\monoid$ over $\State$
which is a monoid with unit $\bot$ and satisfying a few additional
requirements. As a consequence, $\Smonoid$ is a monoidal operation too,
that is essential to combine the realizers of subproofs to obtain a compositional 
interpretation of a proof in terms of interactive realizers. 

The {\em interactive realizability theorem} establishes that for
any theorem $A$ of $\PRA+\EM$ and any interpretation $\alpha$ of
its free variables into individuals, there is a realizer $r(\alpha)$ 
such that $\forces{r(\alpha)}{\alpha}{A}$. As a byproduct
we establish that, given a formula $A(\vec{x},y)$ of $\PRA$ such
that $\PRA+\EM \der A(\vec{x},\phiP(\vec{x}))$, where $\Pred{P}$
is the primitive recursive predicate defined by $A$, we can
extract an effective searching method out of the very proof of $A$, which is 
capable of forcing the interpretation of the free variables in $A$ to values that
satisfy $A$, possibly in some extension of the given state of knowledge,
and actually in the interpretation of $A$ in the standard
classical model of arithmetic. The value computed by $\phiP$ is
such that $A$ holds, but it is not necessarily the best chosen one
w.r.t. all possible usages of $A$. By merging the realizer of $A$ with that
of some proof that uses $A$ as a lemma, we obtain a new realizer
that may resume the search processes of its components, possibly
leading to a new valuation of the $\phiP$. The so obtained
searching method is then an interactive algorithm reflecting
the structure of the proof, that often embodies a cleaver and
efficient idea of how to search for a partial, but locally
sufficient knowledge of the otherwise infinite classical model of
arithmetic.

\section{Primitive recursive arithmetic plus $\EM$ axiom}
\label{sec:primRec}

The theory of primitive recursive arithmetic, called $\PRA$ in
\cite{TroelVD88} (see vol. 1, chapter 3, section 2), is
essentially the quantifier free fragment of Heyting arithmetic
with equality. The language $\LangZero$ of $\PRA$ contains free
variables for natural numbers, the constants $\Fun{0}$ and $\Succ$
for zero and successor respectively; further it includes a
function symbol $\Fun{f},\Fun{g},\ldots$ for each primitive
recursive function, the symbol $=$ for equality and the
connectives $\neg,\wedge,\vee$ and $\rightarrow$. To this list we
add symbols for primitive recursive predicates
$\Pred{P},\Pred{Q},\ldots$ each with a fixed arity.

For presenting $\PRA$ we consider the following deductive system:
the logical axioms are those of $\IPC$, the intuitionistic
predicate calculus, plus the axioms for equality; the non logical
axioms include the defining equations of all primitive recursive
functions and $\neg\,\Succ(\Fun{0}) = \Fun{0}$. As explained in
\cite{TroelVD88}, the formula $\Succ(x) = \Succ(y) \rightarrow x =
y$ is derivable, and needs not to be assumed as an axiom.
Inference rules are:
\[\prooftree A \rightarrow B \quad A\justifies B \using \mbox{\rm MP}\endprooftree \hspace{1cm}
\prooftree A(x) \justifies A(t) \using \mbox{\rm SUB}
\endprooftree \hspace{1cm} \prooftree A(0) \quad A(x) \Then A(\Succ(x)) \justifies
A(y) \using \mbox{\rm IND} \endprooftree \] where in rule
$\mbox{\rm SUB}$ the premise $A(x)$ has been derived from
hypothesis not containing $x$.

By $A(x)$ we mean that $x$ possibly occurs in $A$, and $A(t)$
denotes the same as the more explicit writing $A[t/x]$, namely the
substitution of $t$ for $x$ in $A$. Although there are no bound
variables in the formulas of $\LangZero$, we speak of the sets
$\fv{t}$ and $\fv{A}$ of the free variables occurring in $t$ and
$A$ respectively.

In \cite{TroelVD88} a more general quantifier free version of the
induction rule is considered, namely:
\[\prooftree A(0) \quad A(x) \Then A(\Succ(x)) \justifies
A(t) \endprooftree\] This rule is however admissible by the rules IND
and SUB above, by choosing a fresh $y$ in the conclusion of IND.

The standard model of $\PRA$ interprets terms into $\Nat$, the set
of natural numbers, and function and predicate symbols into their
recursion theoretic counterparts. In the next sections we make use
of the simply typed $\lambda$-calculus with numerals and
recursors, known as G\"{o}del system {\bf T} (see e.g.
\cite{girard:1989a}), for describing our interpretation and
constructions: it is then natural to see the standard model of $\PRA$ inside
the set theoretic model of the typed $\lambda$-calculus, and to
consider predicates as denoting functions with values in the set
$\Bool = \Set{\TrueVal, \FalseVal}$ instead of number theoretic
functions with values in $\Set{0,1}$ as usual. Because of the
absence of quantifiers, it is routine to show that any formula
$A\in\LangZero$ with $\fv{A}\subseteq \Set{x_1,\ldots,x_k}$
defines a $k$-ary primitive recursive predicate.

$\PRA$ is a fragment of the constructive arithmetic $\HA$; however
all the instances of the excluded middle which are expressible in
the language $\LangZero$ are derivable in this theory.

\begin{proposition}\label{prop:PRA_EM} For all $A\in\LangZero$ it is the case that
$\PRA \der A \Or \Not A$.
\end{proposition}

\begin{proof} See \cite{TroelVD88}, Prop. 2.9.
\end{proof}

As a consequence we could freely assume the axioms of $\CPC$, the
classical propositional calculus, in place of $\IPC$. The
essential point here is that the absence of quantifiers makes
excluded middle into an intuitionistically acceptable principle
w.r.t. to an oracle evaluating the function symbols occurring in
$A$. Indeed at the hearth of the proof of Proposition
\ref{prop:PRA_EM} is the fact that we can prove e.g.
$\Fun{f}(\vec{x}) = \Fun{g}(\vec{x})\Or \Fun{f}(\vec{x}) \neq
\Fun{g}(\vec{x})$ by simultaneous induction (which is admissible
in $\PRA$), which is possible only because this formula does not
express that $\Fun{f}$ and $\Fun{g}$ are either equal or different
functions, as this last statement requires the existential
quantification.

\medskip
Let us call $\EM$ the following schema, with $A\in\LangZero$ such
that $\fv{A}\subseteq \vec{x},y$:
\[(\EM)\hspace{8mm} \forall \vec{x}. \;\exists y\; A(\vec{x},y) \Or \forall y\, \Not A(\vec{x},y).\]
$\EM$ is just an instance of the law of excluded middle where
$\exists y\,A(\vec{x},y)$ is a $\Sigma^0_1$ formula with
parameters, and it is called the $\Sigma^0_1$-{\bf LEM} principle
in the hierarchy studied by Akama et alii in \cite{ABHK04}. $\EM$
uses nested quantifiers, hence it is not expressible by a formula
in $\LangZero$. To find a quantifier free equivalent of $\EM$ let
us consider its classically equivalent prenex and skolemized
normal form:
\[\forall \vec{x}, y. \;A(\vec{x},\varphi(\vec{x}) )\Or  \Not A(\vec{x},y),\]
which on turn is (classically) equivalent to
\[\forall \vec{x}, y. \; A(\vec{x},y) \Then A(\vec{x},\varphi(\vec{x}) ).\]
Then we split it into two quantifier free axiom schemata, for
reasons that will be apparent from the technical development in
the subsequent sections of this work:
\[\begin{array}{rl}
(\chi) & \Pred{P}(\vec{x},y)\Then \chiP(\vec{x}) \vspace{2 mm}\\
(\varphi) &  \chiP(\vec{x}) \Then \Pred{P}(\vec{x},\phiP(\vec{x}))
\end{array}\]
where $\Pred{P}$ is a primitive recursive predicate of arity $k+1$
and $\phiP$ and $\chiP$ are a function and a predicate symbol of
arity $k$ respectively. By axioms  $(\chi)$ and $(\varphi)$ the actual meaning of
$\chiP(\vec{x})$ is the predicate $\exists
y.\;\Pred{P}(\vec{x},y)$. Concerning the meaning of $\phiP$ we
note that the derivable implication $\Pred{P}(\vec{x},y)\Then
\Pred{P}(\vec{x},\phiP(\vec{x}))$ is an instance of the {\em
critical axiom} of Hilbert's $\varepsilon$-calculus
\cite{HilbBern70}, writing $\phiP(\vec{x})$ in place of
$\varepsilon y \Pred{P}(\vec{x},y)$, with the restriction that
$\Pred{P}$ has to be primitive recursive. Asking that $\Pred{P}$
is primitive recursive is equivalent to the restriction that the
$A$ in $\EM$ has to be a formula in $\LangZero$, since by the
above remark, these formulas exactly define primitive recursive
predicates.

Let $\LangOne$ be the language of the quantifier free predicate
calculus defined as $\LangZero$ by adding the new symbols $\phiP$
and $\chiP$ to the list of function and predicate symbols
respectively for each predicate symbol $\Pred{P}$ of $\LangZero$:
then $\LangZero\subseteq\LangOne$, and the definitions of free variables and substitution
apply to $\LangOne$ terms and formulas unchanged. Finally, with a slight abuse of
notation, we call $\PRA+\EM$ the theory $\PRA+(\chi)+(\varphi)$,
which is obtained 
by adding all symbols $\chiP$ and $\phiP$ and all
instances of the $(\chi)$ and
$(\varphi)$-axioms to the axioms of $\PRA$.

We finish this preliminary section by observing that $\PRA+\EM$ is
a proper extension of $\PRA$, which can be argued by taking in the
axioms $(\chi)$ and $(\varphi)$ the Kleene predicate as
$\Pred{P}$: the resulting instances are indeed intuitionistically
unacceptable principles. As a matter of fact the $\chiP$
predicates and the $\phiP$ functions are only recursive in the
halting problem. On passing we note that Proposition
\ref{prop:PRA_EM} still holds, namely $\PRA+\EM\der A\vee \neg A$
for any $A\in\LangOne$, and by the same proof.

\section{The state monad and a constructive interpretation of $\PRA+\EM$ formulas}
\label{sec:stateMonad}%

Let $\Pred{R}_0, \Pred{R}_1, \ldots$ be a denumerable list of
predicate symbols in the language of $\PRA$. We assume that it is
an exhaustive enumeration of primitive recursive predicates, in
the sense that $i$ is the G\"odel number of a definition in $\PRA$
of some primitive recursive predicate, associated to $\Pred{R}_i$.
The $\Pred{R}_i$ are called simply predicates, for which we shall
freely use letters $\Pred{P},\Pred{Q},\ldots$ possibly with
indexes. We write $\Pred{P}\equiv\Pred{Q}$ if and only if both
$\Pred{P}$ and $\Pred{Q}$ refer to the same $\Pred{R}_i$, i.e. to
the same syntactical definition of the relative predicate: hence
$\equiv$ is decidable. As said in Section \ref{sec:primRec}, by
the standard model we mean the standard classical interpretation
of $\PRA$, thought seen in the set theoretic model of system {\bf
T}. Except when treating of the semantic interpretation mapping,
we write ambiguously $\Pred{P}(\vec{m},n)$ for $P(\vec{m},n)$,
where $P$ is the primitive recursive predicate interpreting
$\Pred{P}$.

\begin{definition}[States of Knowledge]\label{def:stateKnowledge}
A {\em state of knowledge} (shortly a {\em state}) is a finite
set:
\[s = \Set{\tuple{\Pred{P}_1,\vec{m}_1,n_1},\ldots, \tuple{\Pred{P}_l,\vec{m}_l,n_l}},\]
such that $\Pred{P_1},\ldots,\Pred{P}_l$ are predicate symbols,
and each $\Pred{P}_i$ is a predicate of arity $k_i + 1$, where
$k_i$ is also the length of $\vec{m}_i$, and:
\begin{enumerate}
\item ({\bf model condition}): $\Pred{P}_i(\vec{m}_i,n_i)$ is true in the standard model for all $i=1,\ldots,l$;
\item({\bf consistency condition}):  if $\Pred{P}_i \equiv \Pred{P}_j$ and $\vec{m}_i = \vec{m}_j$ then $n_i = n_j$.
\end{enumerate}
We call $\States$ the set of  states of knowledge.
\end{definition}

A state of knowledge is a finite piece of information about the
standard model of $\PRA$: it says for which tuples of natural
numbers the predicates $\Pred{P}_i$ are known to be true (by the
model condition). The consistency condition implies that in each
state of knowledge $s$ there exists at most one witness $n$ of the
existential statement $\exists y. \Pred{P}(\vec{m},y)$ for each
predicate $\Pred{P}$ and tuple of natural numbers $\vec{m}$. This
$n$ will be the value of $\phiP(\vec{m})$ in the state $s$.

States of knowledge can be presented as a structure $(\States,
\sleq, \bot, \join)$, where $(\States, \sleq)$ is the partial
order defined by $s \sleq s'$ if and only if $s \subseteq s'$; it
has a bottom element $\bot=\emptyset$ and join of compatible
states, $s \join s' = s \union s'$, where $s,s'$ are {\em
compatible}, written $\Compatible{s}{s'}$, if $s,s' \sleq s''$ for
some $s''\in \States$, or equivalently if whenever
$\tuple{\Pred{P},\vec{m},n}\in s$ and
$\tuple{\Pred{P},\vec{m},n'}\in s'$ it is the case that $n = n'$.
$\State$ is also closed under (arbitrary) intersections, and it is
downward closed w.r.t. $\sleq$, namely subset inclusion.

The set $\States$ is decidable, for which it is essential that the
equality $\Pred{P}_i \equiv \Pred{P}_j$ is an identity of
definitions, being the equivalence of primitive recursive
predicates undecidable. By the finiteness of the states
$s\in\States$, the order and the compatibility relations are
computable, as well as the join of two compatible states.

\medskip
The language $\LangOne$ of $\PRA+\EM$
adds to $\LangZero$ the symbols $\chiP$ and $\phiP$ for each predicate symbol
$\Pred{P}$ of $\LangZero$. To interpret the theory $\PRA+\EM$ we
begin by giving meaning to these symbols.

\begin{definition}\label{def:chi-phi-interp}
For each predicate symbol $\Pred{P}$ of arity $k+1$, let
$\Sem{\chiP}:\Nat^k \times \States \rightarrow \Bool$ be defined
by:
\[\Sem{\chiP}(\vec{m},s) = \left\{\begin{array}{ll}
    \TrueVal & \mbox{if $\tuple{\Pred{P},\vec{m},n} \in s$ for some $n$,} \\
    \FalseVal & \mbox{otherwise.}
    \end{array}\right.
\]
Similarly define $\Sem{\phiP}:\Nat^k \times \States \rightarrow
\Nat$ by:
\[\Sem{\phiP}(\vec{m},s) = \left\{\begin{array}{ll}
    n & \mbox{if $\tuple{\Pred{P},\vec{m},n} \in s$ for some $n$,} \\
    0 & \mbox{otherwise.}
    \end{array}\right.
\]
\end{definition}
Because of the consistency condition in Definition
\ref{def:stateKnowledge}, the value of $\Sem{\phiP}(\vec{m},s)$ is
unique. However there exist states $s$ such that
$\Sem{\phiP}(\vec{m},s) \neq \Sem{\phiQ}(\vec{m},s)$ even if
$\Pred{P}$ and $\Pred{Q}$ are equivalent as predicates, though
they have different indexes. In this case $\Pred{P}$ (and its
equivalent $\Pred{Q}$) denotes a non functional predicate; $\phiP$
and $\phiQ$ are also different symbols, that in some models denote distinct
Skolem functions.

Clearly both  $\Sem{\chiP}(\vec{m},s)$ and
$\Sem{\phiP}(\vec{m},s)$ are computable. Note that the
decidability of $\Sem{\chiP}(\vec{m},s)$ makes the default value
$0$ of $\Sem{\phiP}(\vec{m},s)$ effectively distinguishable from
its possible proper value $0$, according to the fact that
$\tuple{\Pred{P},\vec{m},0} \in s$ or not. In any case the meaning
of $\phiP$ is a total computable function.

\begin{lemma}[Monotonicity of $\Sem{\chiP}$]\label{lem:chiMon}
Let $s \sleq s'$, for $s,s'\in \States$:
\begin{enumerate}
\item if $\Sem{\chiP}(\vec{m},s) = \TrueVal$ then $\Sem{\chiP}(\vec{m},s') = \TrueVal$;
\item the inverse implication does not hold in general.
\end{enumerate}
\end{lemma}

\begin{proof}
Immediate: for the second claim take any $\Pred{P},\vec{m},n$ such
that $\Pred{P}(\vec{m},n)$ is true, $s=\bot$,
$s'=\Set{\tuple{\Pred{P},\vec{m},n}}$.
\end{proof}

In order to extend the standard interpretation of $\PRA$ to a
constructive interpretation of $\PRA+\EM$, though in a richer
model, we let the meaning of any term and formula depend on an
extra parameter in $\States$, even if this is essential only when
symbols $\phiP$ or $\chiP$ occur in the term or formula.

To this aim we use a (strong) monad; following \cite{Mog91} we
present monads as Kleisli triples. We quickly summarize the needed
concepts and definitions: see \cite{Mog91} for a treatment of
strong monads and of their use for giving the semantics to
``computational'' types. Let $|\, \Cat|$ be the class of objects of
the category $\Cat$:

\begin{definition}[Kleisli Triple]\label{def:triple}
A {\em Kleisli triple} $(T,\eta,\_^*)$ over a category $\CatC$ is
given by a mapping $T: |\, \Cat| \rightarrow |\, \Cat|$ over the
objects of $\CatC$, a family of morphisms $\eta_X:X\rightarrow
TX\in \CatC$ for each $X\in |\, \Cat|$, and a mapping $\_^*$ such
that $f^*:TX\rightarrow TY\in \CatC$ whenever $f:X\rightarrow
TY\in \CatC$, and the following equations hold:
\begin{enumerate}
\item \label{def:triple-a} $f^* \circ \eta_X = f$,
\item \label{def:triple-b} $\eta_X^* = \Id_{TX}$, 
\item \label{def:triple-c} $g^* \circ f^* = (g^* \circ f)^*$, where $g:Y\arrow TZ\in\CatC$.
\end{enumerate}
\end{definition}
The mapping $\eta_X$ is the {\em inclusion} of $X$ into $TX$;
$f^*$ is called the {\em extension} of the morphism
$f:X\rightarrow TY$ to the morphism $f^*:TX\rightarrow TY$.
A triple $(T,\eta,\_^*)$ defines a
new category, the Kleisli category $\Cat_{T}$, whose class of
objects is $|\,\Cat|$, and morphisms are $\Cat_T[X,Y] =
\Cat[X,TY]$. The identity $\Id_X^{\Cat_T}$ over $X\in |\,\Cat|$ is
$\eta_X\in \Cat_T[X,X] = \Cat[X,TX]$, and composition is given by:
$f\bullet g = f^*\circ g$. Under this reading, the clauses of
Definition \ref{def:triple} imply that $\Cat_T$ is a well defined
category.

\begin{remark}[Monad]\label{rem:monad}{\em Monads arose in
category theory for the study of adjunctions (see e.g.
\cite{McL}), but they have shown to be a fruitful concept also to
treat algebraic structures. A {\em monad} over a category $\Cat$
is a structure $(T,\eta,\mu)$ where $T:\Cat \rightarrow \Cat$ is a
functor, $\eta: \Id_{\Cat} \natTrans T$ and $\mu:T^2 \natTrans T$
(where $T^2=T\circ T$), called the {\em unit} and the {\em
multiplication} of the monad respectively, are natural
transformations such that, for all $X\in|\,\Cat|$:
\[\mu_X\circ \eta_{TX} = \mu_X \circ T\eta_X = \Id_{TX}, ~~~~
\mu_X \circ \mu_{TX} = \mu_X \circ T\mu_X.\] A triple induces a
monad and viceversa (see \cite{Mog91} and the references there).
In fact, to extend the mapping $T: |\, \Cat| \rightarrow |\,
\Cat|$ to a functor it suffices to set $Tf = (\eta_Y\circ f)^*$
for $f:X\rightarrow Y$; w.r.t. such a functor, $\eta$ satisfies
the naturality condition; to get $\mu$ it suffices to set $\mu_X =
\Id_{TX}^*$. Viceversa given the monad $(T,\eta,\mu)$, one
recovers the Kleisli extension by $f^* = \mu_Y\circ Tf$ for
$f:X\rightarrow TY$. Because of this correspondence, we speak of
triples and monads interchangeably.}
\end{remark}

We work in the category of sets $\SetC$, though using only the
part of it which models G\"odel system {\bf T}. Actually we conjecture
that our constructions could be generalized to any ccc with a
natural number object (see \cite{LambekScott}).

We use the simply typed $\lambda$-calculus as metanotation: sets
are denoted as types and morphisms by $\lambda$-terms. By
$X\rightarrow Y$ we denote the object $Y^X$, but sometimes also
the homset $\CatSet[X,Y]$; $X\rightarrow Y \rightarrow Z$
abbreviates $X \rightarrow (Y \rightarrow Z)$, that is the arrow
associates to the right.  Because of the well known isomorphism
$X\times Y\rightarrow Z \simeq X \rightarrow Y \rightarrow Z$, the
same function will be written both in the uncurrified form:
$f(x,y)$ and in the currified one: $f\,x\,y$, according to
convenience; also the more familiar notation $f(x)$ is preferred
to $f\, x$.

\medskip
The following is a monad, which we call the {\em state monad}:

\begin{definition}[The State Monad]\label{def:stateMonad}
We call the tuple $(\SMonad ,\Sunit,\SSext{\_})$ the  {\em state
monad}, where:
\[\begin{array}{rcll}
\SMonad X & = & \State\rightarrow X & \mbox{where $X\in |\,\SetC|$,}\\
\SSunit_X(x) & = & \lambda s\in \State.\,x & \mbox{for $x\in X$,}\\
\SSext{f}(\alpha) & = & \lambda s\in \State.\,f(\alpha(s),s) &
\mbox{for $f:X\rightarrow \SMonad Y \in \SetC$ and $\alpha\in
\SMonad X$.}
\end{array}\]
\end{definition}
Until this will not cause confusion, we shall abbreviate $\SSunit$
by $\Sunit$ and $\SSext{f}$ by $\Sext{f}$.

\begin{proposition}\label{prop:stateMonad}
The tuple $(\SMonad ,\Sunit,\Sext{\_})$ is a Kleisli triple, hence a monad.
\end{proposition}

\begin{proof}
By checking the equations of Definition \ref{def:triple}. For
equation (\ref{def:triple-a}) let $x\in X$ and $f:X\rightarrow
\SMonad Y$, then:
\[(\Sext{f}\circ\Sunit_X)(x) = \Sext{f}(\lambda s.x) = \lambda s'.f((\lambda s.x)(s'),s') =
\lambda s'. f(x,s') = f(x).\] To see (\ref{def:triple-b}), for any
$\alpha\in\SMonad Y$ and $s\in\State$ we have:
\[\Sext{(\Sunit_Y)}(\alpha,s) = \Sunit_Y(\alpha(s),s) = (\lambda\_\,.\alpha(s))(s) = \alpha(s).\]
Eventually to check (\ref{def:triple-c}) let $f$ be as above,
$g:Y\arrow \SMonad Z$, $\alpha\in\SMonad X$ and $s\in\State$.
Then:
\[(\Sext{g}\circ\Sext{f})(\alpha, s)  =
\Sext{g}(\Sext{f}(\alpha),s)
 =  g(\Sext{f}(\alpha,s),s)
 =  g(f(\alpha(s),s),s).\]
On the other hand:
\[\Sext{(\Sext{g}\circ f)}(\alpha,s) = (\Sext{g}\circ
f)(\alpha(s),s) = \Sext{g}(f(\alpha(s)),s) = g(f(\alpha(s),s),s).
\]
\end{proof}

\begin{remark}\label{rem:SMonad-action-morphisms}
{\em According to Remark \ref{rem:monad}, the action of $\SMonad$
over morphisms is:
\[\begin{array}{lll}
\lefteqn{\SMonad f = \Sext{(\Sunit_{Y} \circ f)}} \\
& = & \lambda \alpha \in\SMonad X\ s\in \States. ((\lambda y \in Y
\ s'\in\States. \ y) \circ f)(\alpha (s),s) \\
& = & \lambda \alpha \in\SMonad X\ s\in \States. f(\alpha(s)) \\
& = & \lambda \alpha \in\SMonad X\,.f\circ\alpha,
\end{array}\] for $f:X\arrow Y$. We call $\SMonad f$ the {\em pointwise
extension} of $f$. Note that $\SMonad f = \lambda
\alpha\,.f\circ\alpha$, hence it is just the hom-functor
$\CatSet(\State,-)$.

From Remark \ref{rem:monad} we know that $\mu^{\SMonad}_X =
\Sext{\Id_{\SMonad X}}$, so that by definition unraveling we have
that $\mu^{\SMonad}$ is the diagonalization of its first argument:
\[\mu^{\SMonad}_X(\delta,s) = \delta(s,s), ~~~\mbox{for any $\delta\in\SMonad^2X$ and $s\in\State$.}\]

\if false
For all $s\in\State$ the pair $(\lambda \alpha\in\SMonad
X.\alpha(s),X)$ where $\lambda \alpha\,.\alpha(s):\SMonad X\arrow
X$, is an {\em algebra} of the monad $\SMonad$ (see e.g.
\cite{McL} Definition ...), or $\SMonad$-algebras, which means
that the following equations do hold:
\[\lambda \alpha\,.\alpha(s)\circ\Sunit_X = \Id_X, ~~~~\mbox{and}~~~~
\lambda \alpha\,.\alpha(s)\circ\mu^{\SMonad}_X = \SMonad(\lambda
\alpha\,.\alpha(s))\circ \lambda \alpha\,.\alpha(s).\] For any
$f\in\CatSet[X,Y]$ it is the case that
$f\circ\lambda\alpha\in\SMonad X\,.\alpha(s) =
\lambda\beta\in\SMonad Y\,.\beta(s)\circ \SMonad f$ that is $f$ is
a morphism from $(\lambda\alpha.\alpha(s), X)$ to
$(\lambda\beta.\beta(s), Y)$: categorically speaking this says
that any arrow of $\CatSet$ is a morphism of the category of
$\SMonad$-algebras, known as the Eilenberg-Moore category
$\CatSet^{\SMonad}$. This fact can be rephrased by saying that the
forgetful functor $U:\CatSet^{\SMonad}\arrow\CatSet$ has a left
adjoint which is full and faithful; from our point of view this
simply says that there is a embedding of the standard model of
$\PRA$ into the the model with states at any section ....

For all $s\in\State$ the pair $(\lambda \alpha\in\SMonad
X.\alpha(s),X)$ where $\lambda \alpha\,.\alpha(s):X\arrow\SMonad
X$, is an {\em algebra} of the monad $\SMonad$, or an
$\SMonad$-algebra, which means that the following diagrams
Eilenberg-Moore category $\CatSet^{\SMonad}$.
commute:
\begin {diagram}
X & \rTo^{\Sunit_X} & \SMonad X  & \hspace{3cm} & \SMonad^2 X & \rTo^{\mu_X^{\SMonad} } & \SMonad X\\
   & \rdTo_{\Id_X} & \dTo_{\lambda \alpha\,.\alpha(s)} & & \dTo^{\SMonad(\lambda \alpha\,.\alpha(s))} & & \dTo_{\lambda \alpha\,.\alpha(s)}\\
   &           &  X & & \SMonad X & \rTo_{\lambda \alpha\,.\alpha(s)} & X
\end{diagram}
where $\mu_X^{\SMonad} = \Sext{\Id_{\SMonad X}} = \lambda \delta \in \SMonad^2 X \, \lambda s\in\State\,.\delta(s,s)$. For any $f:X\arrow Y$ we also have that the diagram below commutes:
\begin{diagram}
\SMonad X & \rTo^{\SMonad f} & \SMonad Y \\
 \dTo^{\lambda \alpha\,.\alpha(s)} & & \dTo _{\lambda \beta\,.\beta(s)}\\
X & \rTo_f & Y
\end{diagram}
Hence for all $s\in\State$, any arrow $f\in\CatSet[X,Y]$ is a
morphism of $\SMonad$-algebras $f:(\lambda \alpha\in\SMonad
X.\alpha(s),X)\arrow (\lambda \beta\in\SMonad Y.\beta(s),Y)$,
namely it is an arrow of the Eilenberg-Moore category
$\CatSet^{\SMonad}$.

\fi }\end{remark}

\medskip
Next we extend the definitions of $\Sem{\chiP}$ and $\Sem{\phiP}$
in \ref{def:chi-phi-interp} and we interpret truth values and
numbers expressed by terms and formulas in $\LangOne$ by elements
of $\SMonad\Bool$ and $\SMonad\Nat$ instead of $\Bool$ and $\Nat$
respectively. For the sake of discussion let $t$ and $A$ be any
term and formula of $\LangOne$ respectively such that there is at
most one free variable occurring in them. In case $t\in\LangZero$
its standard interpretation is a map in $\Nat\arrow\Nat$; if
instead $t\in\LangOne\setminus\LangZero$ then its meaning should
have a similar type than $\phiP(x)$ (for some binary $\Pred{P}$)
that is an arrow $\Sem{t}:\Nat\arrow\SMonad\Nat$; similarly the
semantics of $A\in\LangOne\setminus\LangZero$ should be in
$\Nat\arrow\SMonad\Bool$. To define a uniform interpretation of
$\LangOne$ we have two possibilities. The first one is to let
terms and formulas to have their denotations of type
$X\arrow\SMonad Y$, namely as arrows of the Kleisli category
$\CatSet_{\SMonad}$: this is the choice preferred in \cite{Mog91}.
However in our construction we use types of the form $\SMonad X
\arrow \SMonad Y$ for morphisms, because we want to stress that our
``individuals'' are certain convergent objects in $\SMonad X$ for
$X=\Nat$ or $\Bool$ (see Section \ref{sec:convergent}). This is in
analogy with the common idea that real numbers are the actual
individuals of analysis, even if they have been constructed, say,
as Cauchy sequences of rational numbers.

In general given the monad $T$ over the category $\Cat$ we may consider
the category $\CatC_T^*$ such that $|\,\CatC_T^*|= |\,\CatC|$ and
$\CatC_T^*(X,Y) = \Set{f^*\mid f \in\CatC_T(X,Y)}$, where we
recall that $\CatC_T(X,Y)=\CatC(X,TY)$. Then it is straightforward
to see that $\CatC_T^*$ and $\CatC_T$ are equivalent categories,
so that we can work out the interpretation of $\LangOne$ in
$\CatSet_{\SMonad}^*$ without essentially departing from Moggi's
theory of computational types. 

\medskip More in detail we say that an {\em
environment} is a map $\xi:\mbox{\it Var}\rightarrow \SMonad\Nat$
and consistently that the interpretation of a term $t$ of the
language $\LangOne$ should be an element $\Sem{t}^{\SMonad}_{\xi}
\in \SMonad\Nat$. For the basic cases we have the unproblematic
clauses:
\[\Sem{x}^{\SMonad}_{\xi} = \xi(x), ~~~~
\Sem{\sym{0}}^{\SMonad}_{\xi} = \Sunit_{\Nat}(0) = \lambda
\_\,.\,0,\] where we write $\lambda \_\,. \cdots$ for $\lambda s
\in \States.\cdots$ when $\cdots$ does not depend on $s$.

Suppose that $\Fun{f}$ is a unary functional symbol, whose meaning
in the standard model is the (primitive recursive) function
$f:\Nat\rightarrow\Nat$, and that the interpretation
$\Sem{t}^{\SMonad}_{\xi}$ of the term $t$ has been defined; by
taking into account Remark \ref{rem:monad} about the action of
$\SMonad$ over $f$ we can define:
\[\Sem{\Fun{f}(t)}^{\SMonad }_\xi = (\SMonad
f)(\Sem{t}^{\SMonad}_\xi) = \lambda s\in\State.
f(\Sem{t}^{\SMonad}_\xi(s)),\] that is $\Sem{\Fun{f}}^{\SMonad} :=
\SMonad f$. In the case of predicates we have similarly that, if
$P:\Nat\rightarrow \Bool$ is the interpretation of the unary
predicate $\Pred{P}$, then we define $\Sem{\Pred{P}}^{\SMonad} :=
\SMonad P = \lambda \alpha \in\SMonad\Nat\ s\in \States. P(\alpha(
s))$.

\begin{remark}\label{rem:numerals}{\em
The interpretation of a numeral $\sym{n} \equiv \Succ^n\sym{0}$ is
a constant function in $\SMonad\Nat$: if $\mbox{\it
succ}:\Nat\rightarrow\Nat$ is the successor function, then for
example (by omitting the environment $\xi$):
\[\Sem{\Succ (\sym{0})}^{\SMonad}= \SMonad(\mbox{\it
succ})\Sem{\sym{0}}^{\SMonad} = \lambda s \in \States. \ \mbox{\it
succ}\,((\lambda\_\,.0) \ s) = \lambda\_\,. \mbox{\it succ}(0) =
\lambda\_\,.1.
\]
By the interpretation of $\sym{0}$ and a straightforward
induction, we have that $\Sem{\sym{n}}^{\SMonad}=\lambda\_\,. \
n$, for all $n$.}
\end{remark}

\medskip We step to $k$-ary functions and predicates
using the construction proposed in \cite{Mog91}. A uniform
embedding of $TX_1\times \cdots \times TX_k$ into
$T(X_1\times\cdots\times X_k)$ has to be provided, which is
constructed by means of the concept of tensorial strength. If the
reader is not interested to the treatment of product in the theory
of monads and computational types, (s)he may skip the Definition
\ref{def:strongMonad}, the Proposition \ref{prop:tensStrength},
and take the equations (\ref{eq:psi})-(\ref{eq:f-S-interp}) below
\ref{prop:tensStrength} as definitions of tuple and of the
interpretation of $k$-ary functions.

\medskip
Recall that a category has enough points if it has a terminal
object $1$, and for all pair of arrows $f,g:X\arrow Y$, if $f\circ
a = g\circ a$ for all $a:1\arrow X$ (called {\em point} of $X$),
then $f = g$. The category $\SetC$ has obviously enough points
(actually it is the typical such category), as the terminal object
is any singleton set $\Set{*}$, a point $\bar{x}:\Set{*}\arrow X$
is just a constant function $*\mapsto x$ for some $x\in X$ and
morphisms are set theoretic maps; consequently we can use a more compact definition of
tensorial strength than in the general case of cartesian categories, on the ground of Proposition
3.4 of \cite{Mog91}. We also recall that any category theoretic $\lambda$-model has enough
points, so that the following is not really restrictive.

\begin{definition}[Tensorial Strength and Strong Monad \cite{Mog91}]\label{def:strongMonad}
~~Let $(T,\eta,\mu)$ be a monad over a category $\Cat$ with finite
products and enough points, and $!_Y:Y\arrow 1$ the unique
morphism from $Y$ to the terminal object. A {\em tensorial
strength} $t$ of $(T,\eta,\mu)$ is the unique family of morphisms
$t_{X,Y}:X\times TY \arrow T(X\times Y)$ of $\Cat$ such that:
\[\forall a:1\arrow X, b:1\arrow TY.~t_{X,Y}\circ\tuple{a,b} =
T(\tuple{a\circ !_Y,\Id_Y})\circ b.\] If $t$ is a tensorial
strength of $(T,\eta,\mu)$ we say that $(T,\eta,\mu,t)$ is a {\em
strong monad}.
\end{definition}

\begin{proposition}[Tensorial Strength of
$\SMonad$]\label{prop:tensStrength} The state monad $\SMonad$ has
a tensorial strength given by:
\[t_{X,Y}(x, \alpha) \DefEq \lambda s \in \States. \  (x, \alpha(s)),\]
where $(\_\, , \_ )$ is just set theoretic pairing.
\end{proposition}

\begin{proof}
Set $\bar{x}:\Set{*}\arrow X$ to $\lambda\_\,.x$ and
$\bar{\alpha}:\Set{*}\arrow \SMonad Y$ to $\lambda\_\,.\alpha$ for
$x\in X$ and $\alpha\in\SMonad X$, which are the points such that
$(t_{X,Y}\circ\tuple{\bar{x},\bar{\alpha}})(*) = t_{X,Y}(x,
\alpha)$. Now, by Remark \ref{rem:SMonad-action-morphisms} and
since $!_Y = \lambda\_\in Y\,.*$:
\[\begin{array}{lll}
\lefteqn{\SMonad(\tuple{\bar{x}\circ !_Y,\Id_Y}) =
\SMonad(\tuple{\lambda
\_\in Y\,.x,\Id_Y}) }\\
& = &\lambda \beta\in\SMonad Y \ s \in
\State.\tuple{\lambda \_\in Y\,.x,\Id_Y}(\beta(s)) \\
& = & \lambda \beta\in\SMonad Y \ s \in \State.(x,\beta(s)),
\end{array}
\]
which is of type $\SMonad Y\arrow\SMonad(X\times Y)$. Hence :
\[(\SMonad(\tuple{\bar{x}\circ !_Y,\Id_Y})\circ\bar{\alpha})(*) =
(\lambda \beta\in\SMonad Y \ s \in
\State.(x,\beta(s)))(\bar{\alpha}(*)) = \lambda \ s \in
\State.(x,\alpha(s)) = t_{X,Y}(x,\alpha),
\]
and therefore $t_{X,Y}\circ\tuple{\bar{x},\bar{\alpha}} =
\SMonad(\tuple{\bar{x}\circ !_Y,\Id_Y})\circ\bar{\alpha}$ as
desired.
\end{proof}

\noindent Putting $\psi_{X,Y} \DefEq \Sext{(t_{X,Y}\circ
c_{\SMonad Y,X})} \circ t_{\SMonad Y,X} \circ c_{\SMonad X,\SMonad
Y}$, where $c_{X,Y}:X\times Y\arrow Y\times X$ is the canonical
exchange isomorphism, we have a function $\psi_{X,Y}: \SMonad
X\times \SMonad Y \arrow \SMonad (X\times Y)$ which is the
component at $X,Y$ of a natural transformation (see \cite{Mog91},
Remark 3.6). 
By definition unfolding we obtain:
\begin{equation}\label{eq:psi}
\psi_{X,Y}(\alpha, \beta) = \tuple{\alpha,\beta} = \lambda
s\in\State. \ (\alpha(s), \beta(s)).\end{equation} 
This concludes the categorical detour about strong monads.

\medskip
Let $\mbox{\it
eq}:\Nat\times\Nat\rightarrow\Bool$ be the equality map:
$\mbox{\it eq}(m,n) = \TrueVal$ if $m = n$, $\mbox{\it eq}(m,n) =
\FalseVal$ else. If $\Sem{t_1}^{\SMonad}_\xi$ and
$\Sem{t_2}^{\SMonad}_\xi$ have been defined then:
\[(\SMonad (\mbox{\it eq})\circ\psi_{\Nat,\Nat})(\Sem{t_1}^{\SMonad}_\xi,\Sem{t_2}^{\SMonad}_\xi) =
\SMonad (\mbox{\it
eq})\tuple{\Sem{t_1}^{\SMonad}_\xi,\Sem{t_2}^{\SMonad}_\xi} =
\lambda s\in\States. \mbox{\it
eq}(\Sem{t_1}^{\SMonad}_\xi(s),\Sem{t_2}^{\SMonad}_\xi(s)),\]
which is natural to take as the interpretation of equality.

It is straightforward to generalize $\psi$ to $k$-ary products:
\begin{equation}\label{eq:kAry-psi}
\psi_{X_1,\ldots,X_k}:\SMonad X_1 \times \cdots \times
\SMonad X_k
    \arrow \SMonad(X_1 \times \cdots \times X_k)\end{equation}
where, if $\alpha_1\in\SMonad X_1,\ldots,\alpha_k\in\SMonad X_k$
we have:
\[\psi_{X_1,\ldots,X_k}(\alpha_1,\ldots\alpha_k) =
\tuple{\alpha_1,\ldots,\alpha_k}= \lambda s \in
\State.(\alpha_1(s),\ldots,\alpha_k(s)).\] If $f:X_1 \times \cdots
\times X_k \arrow Y$ then we abbreviate:
\begin{equation}\label{eq:f-SMonad}
f^{\SMonad}\DefEq \SMonad(f) \circ \psi_{X_1,\ldots,X_k}: \SMonad
X_1 \times \cdots \times \SMonad X_k \arrow \SMonad
Y,\end{equation} where if $k = 1$ then $f^{\SMonad} \DefEq
\SMonad(f)$.

In particular consider
$\psi_{\Nat,\ldots,\Nat}:(\SMonad\Nat)^k\arrow\SMonad(\Nat^k)$,
with $k$ occurrences of $\Nat$ in the subscript of $\psi$, that we
shall denote shortly by $\psi_{\Nat^k}$. Then we eventually
obtain, for the $k$-ary function symbol $\Fun{f}$ and its
semantics $f:\Nat^k\rightarrow \Nat$ in the standard model:
\begin{equation}\label{eq:f-S-interp}\Sem{\Fun{f}}^{\SMonad } \DefEq f^{\SMonad} =
\lambda \alpha_1 \in \SMonad\Nat\ \ldots \  \alpha_k\in
\SMonad\Nat \ s \in \States. \
f(\alpha_1(s),\ldots,\alpha_k(s)),\end{equation} namely the
semantics of $\Fun{f}$ is the pointwise lifting of its meaning in
the standard model. The same construction works for the $k$-ary
predicates. Finally we abuse notation and write:
\begin{equation}\label{eq:phi-psi-S}
\Sem{\phiP}^{\SMonad}  \DefEq  \Sext{\Sem{\phiP}}\circ
\psi_{\Nat^k} ~~~~\mbox{and} ~~~~ \Sem{\chiP}^{\SMonad}  \DefEq
\Sext{\Sem{\chiP}}\circ \psi_{\Nat^k},\end{equation} where
$\Sem{\phiP}$ and $\Sem{\chiP}$ have been defined in
\ref{def:chi-phi-interp} for all $k+1$-ary predicates $\Pred{P}$ of $\LangZero$, and $\Sext{\_}$
is the extension mapping of the monad $\SMonad$.

\medskip
In summary the interpretation of $\PRA+\EM$ atomic formulas is the
following:
\begin{definition}[Terms and Atomic Formulas Interpretation]\label{def:termAndAtomicFormInterp}
~ Let $\xi:\mbox{\it Var}\rightarrow \SMonad\Nat$ be an
environment for the individual variables, and $t$ a term of the
language $\LangOne$; then $\Sem{t}^{\SMonad}_\xi\in\SMonad\Nat$
is inductively defined:
\[\begin{array}{rcl}
\Sem{x}^{\SMonad }_\xi & = & \xi(x) \vspace{2mm} \\
\Sem{\sym{0}}^{\SMonad }_\xi & = & \Sunit_{\Nat}(0) \vspace{2mm} \\
\Sem{\Fun{f}(t_1,\ldots,t_k)}^{\SMonad }_\xi & = &
\Sem{\Fun{f}}^{\SMonad}(\Sem{t_1}^{\SMonad }_\xi,\ldots,\Sem{t_k}^{\SMonad }_\xi) \vspace{2mm}\\
\Sem{\phiP(t_1,\ldots,t_k)}^{\SMonad }_\xi & = &
\Sem{\phiP}^{\SMonad}(\Sem{t_1}^{\SMonad
}_\xi,\ldots,\Sem{t_k}^{\SMonad }_\xi)
\end{array}\]
where $\Pred{P}$ is a $k+1$-ary predicate symbol.

\medskip\noindent
If $A$ is an atomic formula in $\LangOne$ and $\xi$ an environment
then $\Sem{A}^{\SMonad }_\xi\in\SMonad\Bool$ is defined:
\[\begin{array}{rcl}
\Sem{\Pred{Q}(t_1,\ldots,t_k)}^{\SMonad }_\xi & = &
\Sem{\Pred{Q}}^{\SMonad}(\Sem{t_1}^{\SMonad }_\xi,\ldots,\Sem{t_k}^{\SMonad }_\xi) \vspace{2mm} \\
\Sem{\chiP(t_1,\ldots,t_k)}^{\SMonad }_\xi & = &
\Sem{\chiP}^{\SMonad}(\Sem{t_1}^{\SMonad
}_\xi,\ldots,\Sem{t_k}^{\SMonad }_\xi)
\end{array}\]
where $\Pred{P}$ is a $k+1$-ary predicate symbol.
\end{definition}

\begin{remark}\label{rem:interpretation} {\em By definition unfolding we have:
\[\begin{array}{rcl}
\Sem{\sym{0}}^{\SMonad }_\xi & = & \lambda \_\,. \ 0 \vspace{2mm} \\
\Sem{\Fun{f}(t_1,\ldots,t_k)}^{\SMonad }_\xi & = &
\lambda s \in \States. \ f(\Sem{t_1}^{\SMonad }_\xi(s),\ldots,\Sem{t_k}^{\SMonad }_\xi(s)) \vspace{2mm}\\
\Sem{\phiP(t_1,\ldots,t_k)}^{\SMonad }_\xi & = & \lambda
s\in\States. \ \Sem{\phiP}((\Sem{t_1}^{\SMonad
}_\xi(s),\ldots,\Sem{t_k}^{\SMonad }_\xi(s), \ s)
\end{array}\]
where $f:\Nat^k\rightarrow \Nat$ is the standard interpretation of
$\Fun{f}$, and
\[\begin{array}{rcl}
\Sem{t_1=t_2}^{\SMonad}_\xi & = & \lambda s\in\States. \mbox{\it
eq}(\Sem{t_1}^{\SMonad}_\xi(s),\Sem{t_2}^{\SMonad}_\xi(s))\vspace{2mm}\\
\Sem{\Pred{Q}(t_1,\ldots,t_k)}^{\SMonad }_\xi & = &
\lambda s \in\States. \  Q(\Sem{t_1}^{\SMonad }_\xi(s),\ldots,\Sem{t_k}^{\SMonad }_\xi(s)) \vspace{2mm} \\
\Sem{\chiP(t_1,\ldots,t_k)}^{\SMonad }_\xi & = & \lambda
s\in\States. \ \Sem{\chiP}((\Sem{t_1}^{\SMonad
}_\xi(s),\ldots,\Sem{t_k}^{\SMonad }_\xi(s), \ s)
\end{array}\]
where $Q:\Nat^k\rightarrow \Bool$ is the standard interpretation
of $\Pred{Q}$. Note that in Definition
\ref{def:termAndAtomicFormInterp} we do not need to mention
explicitly the interpretation of the equality, which is actually a
primitive recursive predicate, and hence an instance of
$\Pred{Q}$.}\end{remark}

Let $\NOT:\Bool \arrow \Bool$ be the boolean negation, and $\AAND,
\OOR, \IMPL\Bool\times\Bool\arrow\Bool$ be the respective binary
boolean functions. Then set $\Sem{\STAR}^{\SMonad} \DefEq
\star^{\SMonad}$ for $\star = \NOT, \AAND, \OOR, \IMPL$.

\begin{definition}[Non Atomic Formulas Interpretation]\label{def:FormInterp}
If $A\in\LangOne$ is a non atomic formula and $\xi$ any
environment, then $\Sem{A}^{\SMonad }_\xi\in\SMonad\Bool$ is
defined by cases:
\[\Sem{\neg A}^{\SMonad }_\xi = \Sem{\NOT}^{\SMonad}(\Sem{A}^{\SMonad
}_\xi),~~~~ \Sem{A \star B}^{\SMonad }_\xi =
\Sem{\STAR}^{\SMonad}(\Sem{A}^{\SMonad }_\xi,\Sem{B}^{\SMonad
}_\xi),
\]
for $\star = \wedge, \vee, \rightarrow$.
\end{definition}

\begin{remark}\label{rem:formulas-Kripke-models}{\em
The interpretation of formulas considered above has some
similarities with Kripke semantics of the intuitionistic predicate calculus: in
both cases indeed the meaning of a formula is indexed over a
partial order; more, the states of knowledge can be easily seen as
(finite) possible worlds. However the monotonicity property of
Kripke models fails in our case:
\[ \Sem{A}^{\SMonad }_\xi(s) = \TrueVal \And s \sleq s' \not\Rightarrow  \Sem{A}^{\SMonad }_\xi(s') = \TrueVal.\]
As a counterexample to the implication take $A := \chiP(x) \arrow
(x = \Succ(x))$, where $\Pred{P}(x,y) := x < y$, $s =
\Set{\tuple{\Pred{P},1,2}}$, $s' = \Set{\tuple{\Pred{P},1,2},
\tuple{\Pred{P},0,1}}$ and $\xi(x) = \lambda\_\,.0$. }
\end{remark}

\medskip
The next lemma states a standard property of the interpretations. We
write $\xi[x\mapsto\alpha]$ for the environment whose domain is
$\mbox{\it dom}(\xi)\cup\Set{x}$ and which is everywhere equal to
$\xi$ but in $x$ where it holds $\alpha$.

\begin{lemma}[Substitution Lemma]\label{lem:substitution}
For all $t,t',A \in\LangOne$ and variable $x$:
\[\Sem{t'[t/x]}^{\SMonad}_\xi = \Sem{t'}^{\SMonad}_{\xi[x\mapsto
\Sem{t}^{\SMonad}_\xi]} ~~~\mbox{and}~~~\Sem{A[t/x]}^{\SMonad}_\xi
= \Sem{A}^{\SMonad}_{\xi[x\mapsto \Sem{t}^{\SMonad}_\xi]}.\]
\end{lemma}

\begin{proof}
By induction over $t'$ and $A$.
\end{proof}

\begin{proposition}\label{prop:conservatitvity}
Let $A\in\LangOne$. If $A$ is either a non logical axiom of
$\PRA$, or a logical axiom of $\PRA+\EM$, or an instance of the
$(\varphi)$-axiom, then $\Sem{A}^{\SMonad }_\xi(s) = \TrueVal$ for
any environment $\xi$ and state $s$.
\end{proposition}

\begin{proof} If $t,A\in\LangZero$ let us write $\Sem{t}^{\Nat}_\rho$ and $\Sem{A}^{\Bool}_\rho$
for the respective interpretations of $t$ and $A$ in the standard model w.r.t.
the standard environment $\rho:\mbox{\it Var}\arrow\Nat$.
Then an immediate consequence of the interpretation of symbols in $\LangZero$ by pointwise lifting of their
standard interpretations is that for all environments $\xi:\mbox{\it Var}\arrow\SMonad\Nat$ and $s\in\State$:
\[\Sem{t}^{\SMonad}_\xi(s) = \Sem{t}^{\Nat}_{\rho_\xi} ~~\mbox{and}~~
\Sem{A}^{\SMonad}_\xi(s) = \Sem{A}^{\Bool}_{\rho_\xi}, ~~~~
\mbox{where ${\rho_\xi}(x) = \xi(x,s)$,}\] which can be formally
established by an easy induction over $t$ and $A$.

Now if $A$ is a non logical axiom of $\PRA$ then $A\in\LangZero$
and $\Sem{A}^{\Bool}_{\rho} = \TrueVal$ for any $\rho$, thus
 $\Sem{A}^{\SMonad }_\xi(s) = \Sem{A}^{\Bool}_{\rho_\xi} = \TrueVal$ for any $\xi$ and $s$.

\medskip Let $A\in\LangOne$ be a logical axiom of $\PRA+\EM$. Then there exists an axiom $A'$ of $\IPC$,
the propositional variables $p_1,\ldots,p_k$ and the formulas $A_1,\ldots,A_k\in\LangOne$ such that
$A = A'[A_1/p_1,\ldots,A_k/p_k]$. If $\eta:\mbox{\it PropVar}\arrow\SMonad\Bool$ is an interpretation of the
propositional letters in our model, then by an obvious extension of Lemma \ref{lem:substitution} to
the propositional variables we have:
\[\Sem{A'[A_1/p_1,\ldots,A_k/p_k]}^{\SMonad}_\xi = \Sem{A'}^{\SMonad}_{\xi,\eta} ~~~
 \mbox{where $\eta(p_i,s) =  \Sem{A_i}^{\SMonad}_\xi(s)$}.\]
Then the thesis follows by the fact that $A'$ is a tautology,
since $\IPC$ is a subtheory of $\CPC$, and by reasoning along the same pattern as above.

\medskip
Eventually let $A \DefEq \chiP(\vec{x}) \Then
\Pred{P}(\vec{x},\phiP(\vec{x}))$ be an instance of the
$(\varphi)$-axiom, where $\Pred{P}$ is a primitive recursive
predicate, and let $\xi$ and $s$ be an arbitrary environment and a
state respectively. Then
\[\Sem{A}^{\SMonad }_\xi(s) = \Sem{\IMPL}^{\SMonad}(\Sem{\chiP(\vec{x})}^{\SMonad }_\xi,
\Sem{\Pred{P}(\vec{x},\phiP(\vec{x}))}^{\SMonad }_\xi)(s) =
\Sem{\chiP(\vec{x})}^{\SMonad }_\xi(s) \IMPL
\Sem{\Pred{P}(\vec{x},\phiP(\vec{x}))}^{\SMonad }_\xi(s).\] Now if
$\Sem{\chiP(\vec{x})}^{\SMonad }_\xi(s) = \Sem{\chiP}(\vec{m}, s)
= \FalseVal$, where $\vec{m} = m_1,\ldots,m_k$ for some $k$ and
$m_i = \xi(x_i,s)$ for all $i=1,\ldots,k$, then $\Sem{A}^{\SMonad
}_\xi(s) = \TrueVal$ vacuously. Otherwise
$\tuple{\Pred{P},\vec{m},n}\in s$ for some $n\in\Nat$: this
implies that $\Pred{P}(\vec{m},n)$ is true in the standard
interpretation and that $\Sem{\phiP(\vec{x})}^{\SMonad}_\xi(s) =
\Sem{\phiP}(\vec{m},s) = n$, so that
$\Sem{\Pred{P}(\vec{x},\phiP(\vec{x}))}^{\SMonad }_\xi(s) = \Sem{\Pred{P}(\vec{m},n)} =
\TrueVal$.
\end{proof}

More is actually true, namely that for any $A\in\LangOne$, if
$\PRA+(\varphi)\der A$ then $\Sem{A}^{\SMonad }_\xi(s) = \TrueVal$
for all environment $\xi$ and state $s$: we do not prove this fact
here, since it is a consequence of Theorem \ref{thr:realizability}
(see Corollary \ref{cor:PRA-phi}).

We also observe that Proposition \ref{prop:conservatitvity} fails
in case of the $(\chi)$-axiom. Consider the instance
$\Pred{P}(\vec{x},y)\Then \chiP(\vec{x})$, where
$\Pred{P}(\vec{m},n)$ is true in the standard model for some
$\vec{m},n\in\Nat$. Then there exist infinitely many $s\in\States$
such that $\Sem{\Pred{P}(\vec{x},y)\Then \chiP(\vec{x})}^{\SMonad
}_{[\vec{\alpha},\lambda\_.n/\vec{x},y]}(s) \neq \TrueVal$, for
which it suffices that $\vec{\alpha}(s)=\vec{m}$, but
$\tuple{\Pred{P},\vec{m},n'}$ $\not\in s$ for any $n' \in \Nat$.
Indeed the $(\chi)$-axiom is the essential difference between
$\PRA$ and $\PRA+\EM$.

\section{Convergence, Individuals and Global Functions}
\label{sec:convergent}

In the previous section we have introduced a dynamic (or perhaps epistemic) concept of
individual, which is a map from states of knowledge to individuals
in the ordinary sense. In this section we select a subset of the
maps in $\SMonad \Nat$ and $\SMonad\Bool$ that will represent
(dynamic) individuals and truth values in our model, and show that
the denotation of any term and of any formula in the language of
$\PRA+\EM$ actually is such a kind of map.

\begin{definition}[Sequences, Strong Convergence and Individuals]\label{def:convSeq}
A {\em weakly increasing sequence} over $\States$, shortly a {\em
w.i. sequence}, is some countable subset
$\Set{s_0,s_1,\ldots}\subseteq \States$ such that:
\[s_0 \sleq s_1 \sleq s_2 \sleq \cdots ,\]
that is it is a mapping $\sigma:\Nat\arrow \State$ such that if
$i\leq j$ then $\sigma(i)\sleq \sigma(j)$. Let $\alpha \in \SMonad
X$:
\begin{enumerate}
\item $\alpha\circ\sigma$ is {\em convergent} and it has a {\em limit point}
      $\lim(\alpha\circ\sigma) = x$ if
      \[\exists i \ \forall j. \ (\alpha\circ\sigma)(i) = (\alpha\circ\sigma)(i+j) =
      x;\]
\item $\alpha$ is {\em strongly convergent} if for all w.i. sequences
      $\sigma$, $\alpha\circ\sigma$ is convergent.
\end{enumerate}
We call a strongly convergent $\alpha\in \SMonad X$ an {\em
individual} of $X$.
\end{definition}
When speaking of $\alpha\in\SMonad X$, we use the terms
individual, strongly convergent or just convergent as synonyms.
For each $\alpha\in\SMonad X$ a sequence of states $\sigma$
induces a sequence $\alpha\circ\sigma$ of values in $X$; it has a
limit if it is eventually constant (namely if it becomes stable),
that is we consider the limit w.r.t. the discrete topology over
$X$. Individuals are intended to refer to their limits, although
these are not necessarily unique: in fact the limit of
$\alpha\circ\sigma$ depends on $\sigma$ in general, so that they
can be different for different w.i. sequences.

\begin{definition}[Constant Individuals and Functions with Global State]\label{def:pointwise}
~
\begin{enumerate}
\item $\alpha\in \SMonad X$ is a {\em constant individual} (or just a {\em constant}) if $\alpha=\lambda\_\,.x$, for some $x\in X$;
\item $f:\SMonad X\arrow \SMonad Y$ has {\em global state} if $f (\alpha, s) = f (\lambda\_\,.\alpha(s), s)$,
      for all $\alpha\in \SMonad X$ and $s\in\State$.
\end{enumerate}
\end{definition}
A constant individual is trivially convergent, hence it is an
individual in the sense of Definition \ref{def:convSeq}. Constant individuals correspond to
Kripke's rigid designators (see \cite{Kr:NN}), which are terms
denoting the same object in all possible worlds.

Functions with global state, henceforth called {\em global functions} for short,
can evaluate their functional argument $\alpha$ in the second argument $s$ only: that is
they have essentially  a unique global state, whence the name. In fact  a non global
$f:\SMonad X\arrow \SMonad Y$ is easily constructed by violating this constrain:
let $\alpha\in\SMonad X$ and $h:\State\arrow\State$ be such that
$h(s) = s'$ and $\alpha(s)\neq\alpha(s')$ for certain
$s,s'\in\State$; then the function $f \DefEq \lambda\beta\,.\beta\circ h$ is
not global since $f(\alpha,s) = \alpha(h(s)) = \alpha(s')$, while
$f(\lambda\_\,.\alpha(s),s) = (\lambda\_\,.\alpha(s))(h(s)) =
\alpha(s)$. Note that, if $h$ is strongly convergent, then $f$
sends individuals to individuals, so that the latter property is not sufficient
for a function to be global.

\begin{lemma}[Retraction Lemma]\label{lem:retraction}
Let
 \[\begin{array}{l@{~~~~~ }l}
\Phi: \SMonad(X\arrow Y) \arrow (\SMonad X\arrow\SMonad Y) & \Phi(f)(\alpha,s) \DefEq f(s)(\alpha(s)) \vspace{2mm} \\
\Psi: (\SMonad X\arrow\SMonad Y) \arrow \SMonad(X\arrow Y) &
\Psi(g)(s, x) \DefEq g(\lambda\_. x, s)
\end{array}\]
Then $\Phi\circ\Psi$ is a retraction; moreover the image of $\Phi$
is exactly the set of global functions.
\end{lemma}

\begin{proof}
Let $f\in\SMonad (X\arrow Y)= \State\arrow(X\arrow Y)$; then for
all $s\in \State$ and $x \in X$:
\[(\Psi\circ\Phi)(f)(s,x) = \Phi(f)(\lambda\_.x, s)
= f(s)((\lambda\_.x)(s)) = f(s, x),
\]
therefore $\Psi\circ\Phi = \Id_{\SMonad(X\arrow Y)}$, so that
$(\Phi\circ\Psi)\circ(\Phi\circ\Psi) = \Phi\circ\Psi$ follows. Now
let $\alpha\in\SMonad X = \State\arrow X$ and $s\in \State$; then
we have:
\[
\Phi(f)(\lambda\_.\alpha(s),s) = f(s)((\lambda\_.\alpha(s))(s)) =
f(s)(\alpha(s)) = \Phi(f)(\alpha,s)
\]
that is $\Phi(f)$ is global. On the other hand if $g\in\SMonad X
\arrow \SMonad Y =(\State\arrow X)\arrow(\State\arrow Y)$ is
global and $f=\Psi(g)$ then:
\[
\Phi(f)(\alpha,s) = \Psi(g)(s,\alpha(s)) =
g(\lambda\_.\alpha(s),s) = g(\alpha,s)
\]
that is $(\Phi\circ\Psi)(g) = g$, namely the image of $\Phi$ is
exactly the subset of the global functions in $\SMonad X\arrow
\SMonad Y$.
\end{proof}

As a corollary, global functions are a characterization of ``lifted'' morphisms: 

\begin{corollary}\label{cor:ext-pointwise}
A function $g:\SMonad X\arrow \SMonad Y$ is global if and only if
$g=\Sext{f}$ for some $f:X\arrow \SMonad Y$. Thus $\CatSet_{\SMonad}^*$ is the largest
sub category of $\CatSet_{\SMonad}$ whose arrows are exactly the global functions.
\end{corollary}

\begin{proof} If $f:X\arrow\SMonad Y$ then for all
$\alpha\in\SMonad X$ and $s\in\States$:
\[\Sext{f}(\alpha, s) = f(\alpha(s), s) = f((\lambda\_\,.\alpha(s))(s), s) =
\Sext{f}(\lambda\_\,.\alpha(s), s),\] hence $\Sext{f}$ is global.
Viceversa by Lemma \ref{lem:retraction} if $g$ is global then
$g=\Phi(h)=\lambda\alpha\in\SMonad X \ s\in\States. \ h(s) (\alpha
(s))$, for some $h:\State\arrow(X\arrow Y)$. Set $f(x, s) =
h(s,x)$, so that $f:X\arrow\SMonad Y$; then:
\[g = \lambda\alpha\in\SMonad X \ s\in\States. \ f(\alpha (s), s) = \Sext{f}.\]
\end{proof}

\begin{remark}\label{rem:global}{\em
By Lemma \ref{lem:retraction} global functions in $\SMonad X\arrow
\SMonad Y$ are, up to an application of $\Phi$, families of maps in $X\arrow Y$ indexed
over $\State$. The corollary relates global functions to the
extension map $\Sext{\_}$ of the state monad. To spell it out
further, consider the following:
\[\begin{array}{llll}
F:(X\arrow Y) \arrow (X\arrow\SMonad Y) & \mbox{where} & F(f) &\DefEq \lambda x \in X \ \lambda\, \_ \in \State. f(x) \vspace{2mm}\\
G:(X\arrow\SMonad Y)\arrow\SMonad(X\arrow Y) & \mbox{where} & G(g) &\DefEq \lambda s \in \State \  \lambda x \in X. g(x,s) \vspace{2mm}\\
H:(X\arrow Y) \arrow \SMonad(X\arrow Y) & \mbox{where} & H(f) &\DefEq \lambda\, \_ \in \States. f
\end{array}\]
Then it is easy to see that the following diagram commutes:
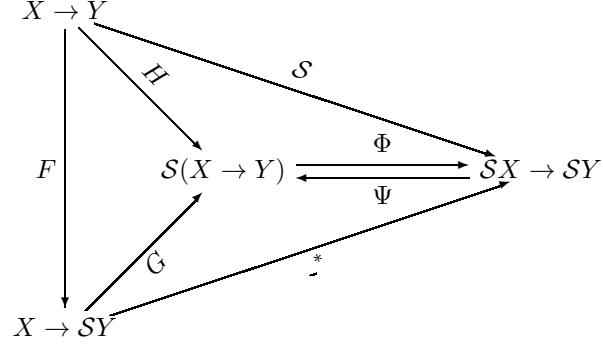
\[\begin{diagram}
X\arrow Y & \\
& \rdTo^H \rdTo(6,2)^{\SMonad} \\
\dTo^F & & \SMonad(X\arrow Y) & &\pile{\rTo^{\Phi}\\ \lTo_{\Psi}} & & \SMonad X\arrow \SMonad Y\\
& \ruTo_G & & & & \ruTo(6,2)_{\Sext{\_}}   \\
X\arrow \SMonad Y
\end{diagram}\]

\medskip
Since $\Phi\circ G = \Sext{\_}$, by Corollary
\ref{cor:ext-pointwise} the image of $G$ is the set of global
maps, up to the embedding $\Phi$. By $\Psi \circ \SMonad = H$, we
see that $H$ is just the action of the functor $\SMonad$ over
arrows, which is nothing more than a pointwise lifting of
functions in $X\arrow Y$ to functions in $\SMonad X\arrow\SMonad
Y$: in Remark \ref{rem:SMonad-action-morphisms} we called the image of $H$ (or more precisely of $\Phi\circ H$)
the subset of {\em pointwise maps} in $\SMonad X\arrow\SMonad Y$.
The fact that $H = G\circ F$ makes it clear that the pointwise
maps are a subset of the global ones. }\end{remark}

The most relevant property of global functions is that their behaviour is
determined by their values over constant individuals.

\begin{theorem}[Density of $\Sunit_X (X)$ in $\SMonad X$]\label{thr:density}
~
\begin{enumerate}
\item \label{thr:density-a}
         If $f,g:\SMonad X\arrow \SMonad Y$ are global and such
         that $f(\lambda\_\,.x) = g(\lambda\_\,.x)$ for all $x\in
         X$, then $f = g$;
\item \label{thr:density-b}
         if $f:\SMonad X\arrow \SMonad Y$ is global and $f(\alpha)$
         is an individual for all constant individuals $\alpha$, then
         $f(\beta)$ is an individual for all individuals $\beta$.
\end{enumerate}
\end{theorem}

\begin{proof}
~
\begin{description}
\item (\ref{thr:density-a}): if $f$ and $g$ are global functions which coincide over constant individuals then
    \[f(\alpha,s) = f(\lambda\_\,.\alpha(s),s) = g(\lambda\_\,.\alpha(s),s) = g(\alpha,s).\]

\item (\ref{thr:density-b}): let $\beta\in\SMonad X$ be an
    individual i.e. strongly
    convergent; then for any w.i. sequence of states $\sigma$ there
    exists $i_0\in\Nat$ such that for all $j\geq i_0$,
    $\beta(\sigma(i_0)) = \beta(\sigma(j))$. Since $f$ is global, we
    know that $f(\beta,s) = f(\lambda\_\,.\beta(s), s)$ for all
    $s\in\State$; therefore
    \[f(\beta,\sigma(j)) = f(\lambda\_\,.\beta(\sigma(j)),
    \sigma(j)) = f(\lambda\_\,.\beta(\sigma(i_0)),
    \sigma(j)),\] for all $j\geq i_0$. By the hypothesis that
    $f(\alpha)$ is strongly convergent for all constant $\alpha$ it follows
    that $f(\lambda\_\,.\beta(\sigma(i_0)))$ is strongly convergent, so that
    there exists $i_1$ such that for all $k\geq i_1$,
    \[f(\lambda\_\,.\beta(\sigma(i_0)),
    \sigma(k)) = f(\lambda\_\,.\beta(\sigma(i_0)),
    \sigma(i_1)).\]
    Then for all $h\geq \max(i_0,i_1)$:
    \[f(\beta,\sigma(h)) = f(\lambda\_\,.\beta(\sigma(i_0)),
    \sigma(h)) = f(\lambda\_\,.\beta(\sigma(i_0)),
    \sigma(i_1)).\]
    We conclude that $f(\beta)$ is strongly convergent.

\end{description}
\end{proof}

Note that $\psi_{X_1,\ldots,X_k}(\alpha_1,\ldots,\alpha_k, s) =
(\alpha_1(s),\ldots,\alpha_k(s))$, so that, if all $\alpha_i$'s are
constant then $\psi_{X_1,\ldots,X_k}(\alpha_1,\ldots,\alpha_k)$ is
such. Strictly speaking (\ref{thr:density-b}) of Theorem \ref{thr:density} does
not apply directly to $\psi$. However this can be proved by a similar and easier argument.

\begin{corollary}\label{cor:psi-convergent}
Each component $\psi_{X,Y}:\SMonad X \times \SMonad Y \arrow
\SMonad (X\times Y)$ of the natural transformation $\psi$ sends
individuals $\alpha,\beta$ into the strongly convergent
$\tuple{\alpha,\beta}=\lambda s\in\State. (\alpha(s),\beta(s))$.
The same holds of $\psi_{X_1,\ldots,X_k}$ for all
$X_1,\ldots,X_k$.
\end{corollary}

\begin{proof}
Given the individuals $\alpha\in\SMonad X$ and $\beta\in\SMonad Y$
and a w.i. sequence $\sigma$ there exist $i_0,i_1\in\Nat$ such
that for all $j$:
\[x = (\alpha\circ\sigma)(i_0+j) = (\alpha\circ\sigma)(i_0) ~~~~ \mbox{and} ~~~~ y = (\beta\circ\sigma)(i_1+j) = (\beta\circ\sigma)(i_1).\]
Therefore for all $i \geq\max(i_0,i_1)$:
\[(\tuple{\alpha,\beta}\circ\sigma)(i) = (\alpha(\sigma(i)), \beta(\sigma(i))) = (x, y).\]
The statement about $\psi_{X_1,\ldots,X_k}$ follows by induction.

\end{proof}

To provide a sufficient condition for the convergence of the output of a map with $k$ arguments,
consider the obvious generalisation of the notion of functions
with global state to the case of $k$-ary functions $f:\SMonad
X_1\times\cdots\times \SMonad X_k \arrow \SMonad Y$, that we call
{\em $k$-global} if for all $\alpha_1\in\SMonad
X_1,\ldots,\alpha_k\in\SMonad X_k$ and $s\in\State$:
\begin{equation}\label{eq:k-global}
f(\alpha_1,\ldots,\alpha_k,s) =
   f(\lambda\_\,.\alpha_1(s),\ldots,\lambda\_\,.\alpha_k(s),s).
\end{equation}

\begin{lemma}\label{lem:k-global}
If $f:\SMonad X_1\times\cdots\times \SMonad X_k \arrow \SMonad Y$
then there exists a unique $\hat{f}:\SMonad(X_1\times\cdots\times
X_k)\arrow \SMonad Y$ such that $f = \hat{f}\circ \psi_{X_1,\ldots,
X_k}$ that is the following diagram commutes:
\begin{diagram}
\SMonad X_1\times\cdots\times \SMonad X_k && \rTo^f && \SMonad Y \\
& \rdTo_{\psi_{X_1,\ldots,
X_k}} && \ruTo_{\hat{f}} \\
&& \SMonad(X_1\times\cdots\times
X_k) &&
\end{diagram}
Moreover $f$ is $k$-global if and only if $\hat{f}$ is global.
\end{lemma}

\begin{proof} Define $\hat{f} \DefEq \lambda\gamma\,.
f(\pi_i\circ\gamma,\ldots,\pi_k\circ\gamma)$.
The first part of the lemma follows by the universal property of the cartesian product.
Indeed we first observe that $\psi_{X_1,\ldots, X_k}$ is a surjective map: if $\gamma\in
\SMonad(X_1\times\cdots\times
X_k)$ and $\pi_i:X_1\times\cdots\times X_k\arrow X_i$ is the $i$-th
projection then:
\[\gamma = \tuple{\pi_1\circ\gamma,\ldots,\pi_k\circ\gamma} = \psi_{X_1,\ldots, X_k}(\pi_1\circ\gamma,\ldots,\pi_k\circ\gamma).\]
Thus, writing $\vec{\alpha}
= \alpha_1,\ldots,\alpha_k$ and $\tuple{\vec{\alpha}} =
\tuple{\alpha_1,\ldots,\alpha_k}$ we know that if $\hat{f}$ exists then:
\[(\hat{f}\circ \psi_{X_1,\ldots, X_k})(\vec{\alpha}) = \hat{f}(\tuple{\vec{\alpha}}) = f(\vec{\alpha}),\]
establishing at the same time unicity and existence of $\hat{f}$.

\noindent
Now if $f$ is $k$-global then for any $s\in\State$:
\[\begin{array}{lll}
\hat{f}(\lambda\_\,.\delta(s),s) & = & f(\pi_1\circ
(\lambda\_\,.\delta(s)), \ldots,
\pi_k\circ(\lambda\_\,.\delta(s)),s) \\
& = & f(\lambda\_\,.(\pi_1\circ \delta)(s), \ldots,
\lambda\_\,.(\pi_k\circ \delta)(s),s) \\
& = & f(\pi_1\circ\delta,\ldots,\pi_k\circ\delta, s) \\
& = & \hat{f}(\delta, s)
\end{array}\]
by the fact that $\pi_i\circ (\lambda\_\,.\delta(s)) =
\lambda\_\,.(\pi_i\circ \delta)(s)$. Viceversa if $\hat{f}$ is global
and $s\in\State$ then:
\[\begin{array}{lll}
f(\vec{\alpha},s) & = & (\hat{f}\circ\psi)(\vec{\alpha},s) \\
& = & \hat{f}(\tuple{\vec{\alpha}}, s) \\
& = & \hat{f}(\lambda\_\,.\tuple{\vec{\alpha}}(s), s) \\
& = & f(\pi_1\circ \lambda\_\,.\tuple{\vec{\alpha}}(s),
\ldots,\pi_k\circ \lambda\_\,.\tuple{\vec{\alpha}}(s)
,s) \\
& = & f(\lambda\_\,.\alpha_1(s),\ldots, \lambda\_\,.\alpha_k(s),
s).
\end{array}\]
\end{proof}

\begin{corollary}\label{cor:k-global}
If $f:\SMonad X_1\times\cdots\times \SMonad X_k \arrow \SMonad Y$
is $k$-global and it sends constant individuals to individuals,
then it sends individuals to individuals.
\end{corollary}

\begin{proof}
Let $\hat{f}$ be the unique global function such that $f = \hat{f}\circ
\psi_{X_1,\ldots, X_k}$, which exists by Lemma \ref{lem:k-global}:
since $\psi_{X_1,\ldots, X_k}$ and its inverse send constant individuals to
constant individuals, $\hat{f}$ sends constant individuals to
individuals by the hypothesis on $f$, so that it sends individuals
to individuals by (\ref{thr:density-b}) of Theorem
\ref{thr:density}. By Corollary \ref{cor:psi-convergent},
$\psi_{X_1,\ldots, X_k}$ also sends individuals to individuals so
that $f$ satisfies the same property.
\end{proof}

\begin{remark}\label{rem:Sf-continuous}{\em If $f:X\arrow Y$ then
by Remark \ref{rem:SMonad-action-morphisms} we have:
\[(\SMonad f)(\lambda\_\,.x) = \lambda s\in\State. f((\lambda\_\,.x)(s))
= \lambda\_\,.f(x),\]
that is $\SMonad f$ sends constants in $\SMonad X$
into constants in $\SMonad Y$.
On the other hand $\SMonad f = \Sext{(\Sunit_X\circ f)}$ is global by
Corollary \ref{cor:ext-pointwise}, so that by Theorem \ref{thr:density}.\ref{thr:density-b},
$\SMonad f$ sends convergent elements into convergent
ones.}\end{remark}

\begin{corollary}\label{cor:k-S-Global}
If $f:X_1\times\cdots\times X_k\arrow Y$ then $f^{\SMonad}$ is
$k$-global. Moreover $f^{\SMonad}$ sends (constant) individuals to
(constant) individuals.
\end{corollary}

\begin{proof}
The first part of the thesis is immediate by Corollary
\ref{cor:ext-pointwise} and Lemma \ref{lem:k-global}. The
remaining part follows 
by $f^{\SMonad} = \SMonad(f) \circ \psi_{X_1,\ldots,X_n}$, the fact that $\SMonad(f)$
sends constant individuals to constant individuals
by Remark \ref{rem:Sf-continuous} and that
the components of the natural transformation $\psi$ send constant individuals to
constant individuals.
\end{proof}

The next lemma relates the interpretation
of terms and formulas to global and $k$-global functions and will be useful in Section \ref{sec:realizTheorem}.

\begin{lemma}\label{lem:k-globalEnv}
For any variable $x$, term $t\in\LangOne$ and formula
$A\in\LangOne$ and for any environment $\xi$ the functions
\[\lambda \alpha\in\SMonad\Nat.
\Sem{t}^{\SMonad}_{\xi[x\mapsto\alpha]}~~~\mbox{and}~~~
\lambda \alpha\in\SMonad\Nat.\Sem{A}^{\SMonad}_{\xi[x\mapsto\alpha]}\] are global. In general
the functions
$\lambda\vec{\alpha}.\Sem{t}^{\SMonad}_{[\vec{\alpha}/\vec{x}]}$
and
$\lambda\vec{\alpha}.\Sem{A}^{\SMonad}_{[\vec{\alpha}/\vec{x}]}$
are $k$-global (for $k$ equal to the length of the vectors
$\vec{\alpha}$ and $\vec{x}$), provided that both $\fv{t}$ and
$\fv{A}$ are included in $\vec{x}$.
\end{lemma}

\begin{proof}
By an easy induction over $t$ and $A$. If $t\equiv x$ then for any $s\in\State$:
\[\Sem{x}^{\SMonad}_{\xi[x\mapsto\alpha]}(s) = \alpha(s) =
(\lambda\_\,.\alpha(s))(s) =
\Sem{x}^{\SMonad}_{\xi[x\mapsto\lambda\_\,.\alpha(s)]}(s).\] The
inductive cases are immediate consequences of the inductive
hypothesis. E.g. let $t\equiv \phiP(t_1,\ldots,t_k)$, then for any
$s\in\State$:
\[\begin{array}{llll}
\Sem{\phiP(t_1,\ldots,t_k)}^{\SMonad}_{\xi[x\mapsto\alpha]}(s) & =
&
\Sem{\phiP}(\Sem{t_1}^{\SMonad}_{\xi[x\mapsto\alpha]},\ldots,\Sem{t_k}^{\SMonad}_{\xi[x\mapsto\alpha]},s)
\\
& = &
\Sem{\phiP}(\Sem{t_1}^{\SMonad}_{\xi[x\mapsto\lambda\_\,.\alpha(s)]},\ldots,\Sem{t_k}^{\SMonad}_{\xi[x\mapsto\lambda\_\,.\alpha(s)]},s)
& \mbox{by ind. hyp.} \\
& = &
\Sem{\phiP(t_1,\ldots,t_k)}^{\SMonad}_{\xi[x\mapsto\lambda\_\,.\alpha(s)]}(s).
\end{array}\]
The rest is equally straightforward.
\end{proof}

An immediate consequence of Remark \ref{rem:Sf-continuous} and
Corollary \ref{cor:k-S-Global} is that if $t$ and $A$ are a term
and a formula in the language $\LangZero$ respectively
(hence not including any $\phiP$ nor $\chiP$ symbol), then their
interpretations $\Sem{t}^{\SMonad}_\xi\in\SMonad\Nat$ and
$\Sem{A}^{\SMonad}_\xi\in\SMonad\Bool$ are constant, provided that each $\xi(x)$ is such. 
We prove now that in general terms and formulas of the language $\LangOne$ 
denote strongly convergent individuals, provided that the free variables occurring in them are interpreted by 
strongly convergent individuals.

\begin{lemma}\label{lem:chiP-phiP-conv}
For all $\vec{m}$ of the appropriate length, both $\lambda
s\in\States.\Sem{\chiP}(\vec{m},s)\in \SMonad\Bool$ and $\lambda
s\in\States.\Sem{\phiP}(\vec{m},s)\in \SMonad\Nat$ are strongly
convergent.
\end{lemma}

\begin{proof}
Consider $\alpha = \lambda s\in\States.\Sem{\chiP}(\vec{m},s)\in
\SMonad\Bool$, and let $\sigma$ be any w.i. sequence. Now either
$\tuple{\Pred{P},\vec{m},n}\not\in\sigma(i)$ for all $i$, so that
$\alpha\circ\sigma$ is the constantly $\FalseVal$ function; or
there exists $i_0$ such that
$\tuple{\Pred{P},\vec{m},n}\in\sigma(i_0)$: then, since $\sigma$
is weakly increasing, $\tuple{\Pred{P},\vec{m},n}\in\sigma(j)$ and
$\alpha(\sigma(j)) = \TrueVal$ for all $j\geq i_0$.

The case of $\lambda s\in\States.\Sem{\phiP}(\vec{m},s)\in
\SMonad\Nat$ is similar.
\end{proof}

\begin{theorem}\label{thr:convInterp}
If $t$ is any term and $A$ any formula of the language
$\LangOne$ and $\xi$ an environment whose domain includes the free
variables of $t$ and $A$, such that $\xi(x)$ is strongly
convergent for all $x$, then both
$\Sem{t}^{\SMonad}_\xi\in\SMonad\Nat$ and
$\Sem{A}^{\SMonad}_\xi\in\SMonad\Bool$ are strongly convergent.
\end{theorem}

\begin{proof} By induction over $t$ and $A$, using
Lemma \ref{lem:chiP-phiP-conv} for the cases
$\phiP(t_1,\ldots,t_k)$ and $\chiP(t_1,\ldots,t_k)$ respectively,
and Corollary \ref{cor:k-S-Global} for the remaining inductive steps.
\end{proof}

\section{The realizers monad}
\label{sec:interactiveReal}

The compositional construction of the search of a state that makes true a certain formula,
which we shall describe in Section \ref{sec:realizTheorem},
rests on the ability of merging pairs of states, even if incompatible.
We give an abstract definition of a merge operation $\monoid$, and show that there exists one such.
In fact more concrete and non-equivalent definitions of merge are possible, as we suggest in a remark.

We then define a quadruple $(\RMonad,\Runit,\Rext{\_}, \monoid)$, parametric in the merge operation $\monoid$,
such that $\RMonad X = \SMonad (X) \times \SMonad(\State)$
is the type of pairs of an individual and a realizer (a concept defined in the next section) interacting each other to the extent of
satisfying a formula, which is the goal of the interaction.

The monoidal structure of the merge is lifted to the maps in $\SMonad(\State)$, 
in order for to meet the requirement for the functor $\RMonad$ to be a monad.

\begin{definition}[Merge]\label{def:mergeM}
A {\em merge} is a mapping
$\monoid:\State\times\State\arrow\State$ such that, for all $s_1,
s_2\in\State$:
\begin{enumerate}
\item \label{def:mergeM-a} $(\State,\monoid,\bot)$ is a monoid;
\item \label{def:mergeM-b} if $s_1\monoid s_2 = \bot$ then $s_1 = \bot = s_2$;
\item \label{def:mergeM-c} $s_1\monoid s_2\subseteq s_1\union s_2$.
\end{enumerate}
\end{definition}

Note that in clause (\ref{def:mergeM-c}) above we cannot write
$s_1\monoid s_2\sleq s_1\join s_2$ since $s_1\union s_2$ might be
an inconsistent set, so not in $\State$.

\begin{lemma}\label{lem:mergeMprop}
If $\monoid$ is a merge then for all $s, s_1, s_2 \in\State$:
\begin{enumerate}
\item \label{lem:mergeMprop-a} if $\Compatible{s}{s_1}$ and $\Compatible{s}{s_2}$ then
    $\Compatible{s}{(s_1\monoid s_2)}$;
\item \label{lem:mergeMprop-b} if $s \cap s_1 = s \cap s_2 = \bot$ then
    $s \cap (s_1\monoid s_2) = \bot$.
\end{enumerate}
\end{lemma}

\begin{proof}
\begin{description}
\item (\ref{lem:mergeMprop-a}): $s\not\uparrow (s_1\monoid s_2)$
    implies that there exists $a\in s_1\monoid s_2$ such that $s
    \not\uparrow \Set{a}$. Since $s_1\monoid s_2\subseteq s_1\union
    s_2$,
    it is the case that $a \in s_i$ for either $i = 1$ or $i = 2$,
    contradicting $\Compatible{s}{s_1}$ and $\Compatible{s}{s_2}$.
\item (\ref{lem:mergeMprop-b}): we observe that
    $s \cap (s_1\monoid s_2) \subseteq s \cap (s_1 \cup s_2) =
    (s\cap s_1)\cup(s\cap s_2)$,
    hence if $(s\cap s_1)\cup(s\cap s_2) = \bot = \emptyset$ then
    $s \cap (s_1\monoid s_2) = \bot$.
\end{description}
\end{proof}

A very simple example of ``merging'' consists in dropping one of the merged states.

\begin{proposition}
The following mapping is a merge:
\[s_1\monoid_0 s_2 = \left\{\begin{array}{ll}
        s_1 & \mbox{if $s_1 \neq \bot$,} \\
        s_2 & \mbox{otherwise.}
        \end{array}\right.
\]
\end{proposition}

\begin{proof} It is immediate that $s_1 \monoid_0 s_2 \in \State$
    for all $s_1, s_2 \in \State$.

    For all $s\in \State$, $\bot\monoid_0 s = s$. If $s = \bot$ then $s\monoid_0\bot = \bot =s$. On the other
    hand if $s \neq \bot$ then $s\monoid_0 \bot = s$: hence $\bot$ is the unit of
    $\monoid_0$.

    Let $s_1 = \bot$, then
    \[(s_1\monoid_0 s_2) \monoid_0 s_3 = s_2 \monoid_0 s_3 = s_1 \monoid_0
    (s_2\monoid_0 s_3).\]
    Else, if $s_1 \neq \bot$ then
    \[(s_1\monoid_0 s_2) \monoid_0 s_3 = s_1 \monoid_0 s_3 = s_1 = s_1 \monoid_0
    (s_2\monoid_0 s_3).\]
    Therefore (\ref{def:mergeM-a}) of Definition \ref{def:mergeM}
    holds.

    If $s_1 \neq \bot$ then $s_1\monoid_0 s_2 = s_1 \neq \bot$;
    similarly if $s_1 = \bot$ and $s_2 \neq \bot$ then $s_1\monoid_0 s_2 = s_2 \neq
    \bot$, so that condition (\ref{def:mergeM-b}) of Definition \ref{def:mergeM}
    follows by contraposition.

    Finally $s_1\monoid_0 s_2 = s_i$ for either $i=1,2$, hence
    (\ref{def:mergeM-c}) of Definition \ref{def:mergeM} is satisfied.
\end{proof}

\begin{remark}\label{rem:merge}
{\em The map $\monoid_0$ is essentially a selector of non $\bot$-states, with a bias toward its first argument: it considers the
second argument just in case the first one is not informative at
all. In particular it is not commutative, while it is clearly
idempotent: $s \monoid_0 s = s$. It is a very simple, thought crude example of merge.
Beside it and its symmetric $s_1 \monoid_0' s_2 \DefEq s_2
\monoid_0 s_1$, there exist other examples of merge that one could
consider. We mention two of them omitting proofs.
\begin{itemize}
\item A ``parallel'' non-commutative merge. Define $\bar{s} = \Set{\tuple{\Pred{P},\vec{m},n}\mid \exists n'. \tuple{\Pred{P},\vec{m},n'} \in s}$,
    and set \[s_1 \monoid_1 s_2 \DefEq s_1 \cup (s_2\setminus
    \bar{s}_1).\]
    This merge saves all of the information in $s_2$ which is consistent
    with $s_1$, while in case of inconsistency, the elements of $s_1$ prevail:
    hence it is not commutative, and its symmetric is a different
    merge. This is the merge operation used in \cite{AschieriB09}.
\item A ``parallel'' commutative merge. For any $X\subseteq\bigcup\State$ define
    $\widehat{X} \DefEq \Set{\tuple{\Pred{P},\vec{m},n} \in X \mid  \forall \tuple{\Pred{P},\vec{m},n'} \in X. n \leq n' }$. Then we set:
    \[ s_1 \monoid_2 s_2 \DefEq \widehat{s_1 \cup s_2}.\]
    The effect of $\widehat{X}$ is, for all
    predicate $\Pred{P}$ and vector of numbers $\vec{m}$, to select, among all possibly inconsistent tuples
    $\tuple{\Pred{P},\vec{m},n_1},\tuple{\Pred{P},\vec{m},n_2},\ldots$ in $X$, the tuple
    $\tuple{\Pred{P},\vec{m},n_i}$, where $n_i$ is the minimum among $n_1, n_2, \ldots$. It follows that
     $\widehat{X}$ is always consistent and, if $X\subseteq \State$ is finite, then it is an element of $\State$. Moreover
     $\widehat{X} \subseteq X$ and $\widehat{\widehat{X}} = \widehat{X}$, hence it is an interior operator. The
     remarkable property of $\monoid_2$ is commutativity. This merge appears in \cite{Berardi05}.
\end{itemize}
We observe that $\monoid_0, \monoid_1$ and $\monoid_2$ are all
computable functions.}
\end{remark}

\bigskip
Since a merge is a function in $\monoid:
\State\times\State\arrow\State$ it can be pointwise lifted to the
mapping $\monoid^{\SMonad} =
\SMonad(\monoid)\circ\psi_{\State,\State}:\SMonad(\State)\times\SMonad(\State)
\arrow\SMonad(\State)$, where $(r\Smonoid r')(s) = r(s) \monoid r'(s)$. By means of $\Smonoid$ we may define a new monad:

\begin{definition}[The Realizer Monad]\label{def:realMonad}
Let $\monoid$ be a merge. Then we say that the tuple $(\RMonad,
\Runit, \Rext{\_},\monoid)$ is a {\em realizer monad} if:
\[\begin{array}{rcll}
\RMonad X & = & \SMonad(X) \times \SMonad(\State) & \mbox{where $X\in |\,\SetC|$,}\vspace{2mm} \\
\Runit_X(x) & = & (\SSunit_X(x), \SSunit_{\State}(\bot)) & \mbox{for $x\in X$,} \vspace{2mm} \\
\Rext{f}(\alpha,r) & = & (\SSext{f}_1(\alpha), r \monoid^{\SMonad}
\SSext{f}_2(\alpha)) & \mbox{for $f:X\rightarrow \RMonad Y \in
\SetC$ and $(\alpha,r)\in \RMonad X$,}
\end{array}\]
where $f_i = \pi_i\circ f$, for $i = 1,2$.
\end{definition}
Below we write $\Sunit$ and $\Sext{\cdot}$ for $\SSunit,\SSext{\cdot}$ respectively, to simplify the notation, while keeping
$\Runit,\Rext{\cdot}$ to distinguish the unit and the extension map of the monad $\RMonad$.

The set $\RMonad X = \SMonad(X) \times \SMonad(\State) $ is larger than its part of interest: as shown in Section \ref{sec:convergent},
the relevant part of $\SMonad(X)$ is the set of individuals; on the other hand, as we shall see in the next section, we
concentrate on realizers which are individuals in $\SMonad(\State)$ satisfying some further conditions. So that there is a slight  abuse of terminology.
However the monad provides an elegant way of pairing individuals and transformations over the states, which is at
the basis of the forcing relation and the realizability interpretation we shall meet in Section \ref{sec:realizTheorem}.

A realizer monad is built on top of the state monad
$(\SMonad,\SSunit,\SSext{\cdot})$, and it is parametric in the merge
$\monoid$. By definition unfolding we have:
\[\RMonad X = (\State\arrow X) \times (\State\arrow \State),
~~~~~~~~ \Runit_X(x) = (\lambda \_ . x, \lambda \_ . \bot),\] and
\[r \monoid^{\SMonad} \Sext{f}_2(\alpha) = \lambda
s\in\State.\,r(s) \monoid f_2(\alpha(s),s).\] 
To better understand
the definition of $\Rext{f}$ observe that the function
\[f:X\arrow [(\State\arrow Y)\times(\State\arrow \State)]\] is
identified with the pair $\tuple{f_1,f_2}$ (as they are the same
 in any cartesian category), so that:
\[f_1:X\arrow (\State\arrow Y) ~~~~\mbox{and}~~~~ f_2:X\arrow (\State\arrow \State),\]
and therefore
\[\Sext{f}_1:(\State\arrow X) \arrow (\State\arrow Y) ~~~~\mbox{and}~~~~
\Sext{f}_2:(\State\arrow X) \arrow (\State\arrow \State).\]
Of these the component $\Sext{f}_1$ is intended to associate individuals over $X$ to individuals over $Y$;
the second and more relevant component $\Sext{f}_2$ formalises how functions over $\State$, in particular realizers, 
can depend on individuals. In particular the importance of merging of $r$ with $\SSext{f}_2(\alpha)$ as the second component of $\Rext{f}(\alpha,r)$ will be discussed in Remark \ref{rem:isomorphism}.

\begin{lemma}\label{lem:S-monoid}
If $\monoid$ is a merge, then $(\SMonad(\State),
\monoid^{\SMonad}, \lambda\_\,.\bot)$ is a monoid.
\end{lemma}

\begin{proof} Observe that, by Remark \ref{rem:SMonad-action-morphisms}
and the definition of $\psi$, $(r\monoid^{\SMonad} r')(s) = r(s)
\monoid r'(s)$ for all $s\in\State$, so that the fact that
$\monoid$ is a monoidal operation over $\State$ with unit $\bot$
immediately implies that $(\SMonad(\State), \monoid^{\SMonad},
\lambda\_\,.\bot)$ is a monoid: for example $(r\monoid^{\SMonad}
\lambda\_\,.\bot)(s) = r(s) \monoid \bot = r(s)$ for all $s$, so
that $r\monoid^{\SMonad} \lambda\_\,.\bot = r$.
\end{proof}



\begin{theorem}\label{thr:realMonad}
If $(\RMonad, \Runit, \Rext{\_},\monoid)$ is a realizer monad then
$(\RMonad, \Runit, \Rext{\_})$ is a Kleisli triple, and hence a
monad.
\end{theorem}

\begin{proof} By checking that $(\RMonad, \Runit, \Rext{\_})$ satisfies the
three equations of Definition \ref{def:triple}, and using the fact
that, by Lemma \ref{lem:S-monoid}, if $\monoid$ is a merge then
$\Smonoid$ is a monoidal operation over $\SMonad(\State)$ with
unit $\lambda\_ \,. \bot$.
\begin{description}
\item $\Rext{f} \circ \Runit_X = f$, that is the following diagram
commutes:
\[\begin{diagram}
X & \hspace{2cm} & \\
\dTo^{\Runit_X} & \rdTo^{f=\tuple{f_1,f_2}}& \\
(\State\arrow X)\times (\State\arrow \State) & \rTo_{\Rext{f}} &
(\State\arrow Y)\times (\State\arrow \State)
\end{diagram}\]
Given $x\in X$ we have:
\[ (\Rext{f}\circ \Runit_X)(x)  =  \Rext{f}(\lambda \_. x, \lambda\_ . \bot) =  ( \Sext{f}_1(\lambda \_. x), \lambda\_ . \bot\monoid^{\SMonad}\Sext{f}_2(\lambda \_. x))
= ( \Sext{f}_1(\lambda \_. x), \Sext{f}_2(\lambda \_. x)), \]
since $\lambda\_ . \bot$ is the unit of $\Smonoid$. But for all
$s\in \State$, $\Sext{f}_1(\lambda \_. x, s) = f_1(x)$, and
$\Sext{f}_2(\lambda \_. x, s) = f_2(x)$, and therefore we get:
\[  (\Rext{f}\circ \Runit_X)(x)  = (f_1(x), f_2(x)) = \tuple{f_1,f_2}(x) =  f(x).\]

\item $\Rext{(\Runit_Y)} = \Id_{\RMonad Y}$.
    By definition $\Runit_Y=\tuple{\Sunit_Y,\lambda \_\in Y.\, \Sunit_{\State}(\bot)}$, where
    for any $\alpha\in\SMonad Y$ we have:
    \[\Sext{(\lambda \_\in Y.\, \Sunit_{\State}(\bot))}(\alpha) = \lambda s\in\State\,.(\lambda\_\,.\bot)(s) = \lambda\_\,.\bot\]
    Now for all $r:\State\arrow \State$ we have:
    \[\Rext{(\Runit_Y)}(\alpha, r) = (\Sext{(\Sunit_Y)}(\alpha), r\Smonoid \lambda\_\,.\bot) = (\alpha,r),\]
    by the fact that $\Sext{(\Sunit_Y)} = \Id_{\SMonad Y}$
    (Proposition \ref{prop:stateMonad}) and that $\lambda\_\,.\bot$ is the unit of
    $\Smonoid$.

\item $\Rext{g} \circ \Rext{f} = \Rext{(\Rext{g} \circ f)}$, where
    $g:Y\arrow\RMonad{Z}$: let again $f = \tuple{f_1,f_2}$ and $g = \tuple{g_1,g_2}$;
     given $\alpha:\State\arrow X$ and $r:\State\arrow \State$, by definition unfolding we have:
    \[\begin{array}{lll}
      (\Rext{g} \circ \Rext{f})(\alpha,r) & = & \Rext{g}(\Sext{f_1}(\alpha), r\Smonoid \Sext{f_2}(\alpha)) \\
      & = & ((\Sext{g_1}\circ\Sext{f_1})(\alpha), (r\Smonoid \Sext{f_2}(\alpha)) \Smonoid (\Sext{g_2}\circ\Sext{f_1})(\alpha))\\
    \end{array}\]
    On the other hand let $h = \Rext{g} \circ f = \tuple{h_1,h_2}$, where $h_i = \pi_i\circ\Rext{g} \circ f$. Then
    \[\Rext{(\Rext{g} \circ f)}(\alpha, r) = (\Sext{h_1}(\alpha), r\Smonoid \Sext{h_2}(\alpha)).\]
    Now:
    \[\begin{array}{lll}
    \Sext{h_1}(\alpha) & = & \lambda s\in\State\,.(\pi_1\circ\Rext{g} \circ f)(\alpha(s),s) \\
    & = & \lambda s\in\State\,. \pi_1((\Rext{g} \circ f)(\alpha(s))) (s) \\
    & = & \lambda s\in\State\,. \pi_1(\Rext{g}(f_1(\alpha(s)),f_2(\alpha(s)))) (s)\\
    & = & \lambda s\in\State\,.(\Sext{g_1}( f_1(\alpha(s)))) (s) \\
    & = & \lambda s\in\State\,.(\Sext{g_1}\circ f_1)(\alpha(s), s) \\
    & = & \Sext{(\Sext{g_1}\circ f_1)}(\alpha).
    \end{array}\]
    Similarly we have:
    \[\begin{array}{llll}
    \Sext{h_2}(\alpha) & = & \lambda s\in\State\,.(\pi_2\circ\Rext{g} \circ f)(\alpha(s),s) \\
    & = & \lambda s\in\State\,. \pi_2((\Rext{g} \circ f)(\alpha(s))) (s) \\
    & = & \lambda s\in\State\,. f_2(\alpha(s))\Smonoid (\Sext{g_2}\circ f_1)(\alpha(s)) (s) \\
    & = & \Sext{f_2}(\alpha) \Smonoid  \Sext{(\Sext{g_2}\circ f_1)}(\alpha) & \mbox{by claim (\ref{eq:mergeGlobal}) below,}
    \end{array}\]
    where
    \begin{equation}\label{eq:mergeGlobal}
    (\Sext{f}(\alpha) \monoid^{\SMonad} \Sext{g}(\alpha))(s) =
    (f(\alpha(s))\monoid^{\SMonad} g(\alpha(s)))(s)
    \end{equation}
    is easily checked by definition unfolding.
    Summing up:
    \[\Rext{(\Rext{g} \circ f)}(\alpha, r) =
    (\Sext{(\Sext{g_1}\circ f_1)}(\alpha), r \Smonoid (\Sext{f_2}(\alpha) \Smonoid  \Sext{(\Sext{g_2}\circ f_1)}(\alpha))),\]
    and we conclude by noting that $\Sext{(\Sext{g_1}\circ f_1)} = \Sext{g_1}\circ\Sext{f_1}$,
    $\Sext{(\Sext{g_2}\circ f_1)} = \Sext{g_2}\circ\Sext{f_1}$ since $\Sext{\_}$ is the extension of a Kleisli triple
    by Proposition \ref{prop:stateMonad}, and because $\Smonoid$ is associative.

\end{description}
\end{proof}

\begin{remark}\label{rem:isomorphism}
{\em In any ccc it is the case that
\[(Z\arrow X)\times (Z\arrow Y) \simeq Z\arrow (X\times Y)\]
is a natural isomorphism, given by $(f,g) \mapsto \tuple{f,g}$.
Therefore
\[  \RMonad X = (\State\arrow X) \times (\State\arrow \State) \simeq \State \arrow (X\times \State),\]
which is equal to $\SMonad (X\times \State)$. $(\RMonad, \Runit, \Rext{\_})$ is similar to the
side effect monad $(\EMonad, \Eunit, \Eext{\_})$ in \cite{Mog91}:
 \[\begin{array}{lll}
           \EMonad X & = & S\rightarrow(X\times S) \vspace{1mm} \\
           \Eunit_X(x) & = & \lambda s \in S.(x,s) = \tuple{\Sunit_X(x),\Id_S} \vspace{1mm} \\
           \Eext{f}(\gamma) & = & \lambda s \in S. \ f((\pi_1\circ \gamma)(s), (\pi_2\circ
           \gamma)(s))
      \end{array}\]
where $S$ is some set of states, $f:X\arrow (S\arrow(Y\times S))$
and $\gamma:S\arrow(X\times S)$. In case of $S = \State$ we have
that if $(\alpha,r)$ is a pair of a convergent mappings, then
$\tuple{\alpha,r}:S\arrow(X\times \State)$ is such (see Corollary
\ref{cor:psi-convergent}): therefore the isomorphism $\RMonad X
\simeq \SMonad (X\times \State)$ preserves convergence.

The computational idea behind $\EMonad$ and $\RMonad$ is however
different. In the case of the side effects monad the function
$\Eext{f}(\gamma)(s)$, where $\gamma = \tuple{\gamma_1,\gamma_2}$,
first evaluates $\gamma_1$ in the state $s$, possibly leading to a
new state $s'=\gamma_2(s)$, intuitively because of side effects in
the evaluation of $\gamma_1(s)$; then
$f(\gamma_1(s)):S\arrow(Y\times S)$ is evaluated in the new state $s'$.
This is necessarily a sequential process.

In the case of the realizer monad the function $\Rext{f}(\alpha,r) = (\Sext{f}_1(\alpha), r\Smonoid
\Sext{f}_2(\alpha))$ first computes a new dynamic object $\Sext{f}_1(\alpha)$, then forces it to satisfy some property using the realizer $\Sext{f}_2(\alpha)$ merged with the realizer $r$, that is supposed to satisfy some other (possibly different) property.
The reason is that the search procedure represented by  $\Sext{f}_2(\alpha)$ might change the state reached by some previous attempt by $r$ to force $\alpha$ into its own goal, destroying the work by $r$. Hence both $\Sext{f}_2(\alpha)$ and $r$ have to be kept while evaluating the realizer obtained by their merge, and cannot be sequentialized.
}\end{remark}

\section{Interactive realizers}\label{sec:realizTheorem}

This section introduces the central concepts of {\em interactive
realizer} and of {\em interactive forcing}, which are the main
contribution of our work. Realizers have been introduced by Kleene
as an interpretation of Brouwer's and Heyting's concept of
construction. In the case of constructive theories a realizer is a
direct computation, possibly depending on some parameters. With a
non constructive theory like $\PRA+\EM$ the saving of such an idea
involves the shift from recursiveness to recursiveness in the
limit. In this perspective a realizer is not an algorithm (a
recursive function), rather it is the recursive generator of a
search procedure that, along a series of attempts and failures,
eventually attains its goal.

\begin{definition}[Interactive Realizers]\label{def:realizer}
An {\em interactive realizer} is a map $r\in\SMonad(\State) =
\State\arrow\State$, such that:
\begin{enumerate}
\item\label{def:realizer-a} $r$ is strongly convergent;
\item\label{def:realizer-b} $r$ is  {\em compatible with its arguments} that is:
$\Compatible{r(s)}{s}$ for all $s\in \State$;
\item\label{def:realizer-c} $r(s) \cap s = \bot$ for all $s\in\State$.
\end{enumerate}
A state $s\in\State$ is a {\em prefix point} of
$r:\State\arrow\State$ if $r(s) \sleq s$; by $\Prefix{r}$ we
denote the set of prefix points of $r$.
\end{definition}

\begin{remark}\label{rek:realizerDef}
{\em By clause (\ref{def:realizer-a}) above the realizers are
individuals over $\State$. Note that identity over $\State$ is not
convergent, and so it is not a realizer. Compatibility condition
(\ref{def:realizer-b}) is essential, together with convergence,
for the existence of pre-fixed points: see Proposition
\ref{prop:cofin} below. The function $\lambda \_\,.\bot$ is a
(trivial) realizer and, because of clause (\ref{def:realizer-c}),
the only one among constant individuals.

By clause (\ref{def:realizer-c}), if $r$ is a realizer we have
that $s\in\Prefix{r}$ if and only if $r(s) = \bot$, because $r(s) \sleq s$ implies
$r(s) \cap s = \bot$.
Namely $\Prefix{s}$ is the set of ``roots'' of $r$.
This clause is just intended to simplify the treatment of realizers, in the sense that
if $r$ is a realizer,  then $r(s)$ just adds ``new'' atoms to $s$; hence if $r(s) \sleq s$
this means that there is actually nothing to add.

\if false
To explain the point
let us consider the following equivalence over functions in
$\SMonad(\State)$ which are compatible with their arguments
(recall that the stales are sets and that $\State$ is closed w.r.t.
$\supseteq$):
\[f \approx g \Iff \forall s.~ f(s)\setminus s = g(s)\setminus
s.\]
It is easy to check that $f \approx g$ implies $\Prefix{f}=\Prefix{g}$.
The intuitive meaning of $f \approx g$ is that they add the
same ``new'' tokens w.r.t. their arguments. As a consequence $f
\approx f \cup \Id_{\State} \approx f \setminus \Id_{\State}$ for
all compatible $f$, where union and set difference are pointwise,
and the former is well defined by the compatibility of $f$ with its argument. Now
clause (\ref{def:realizer-c}) has the effect to choose $f'\DefEq f
\setminus \Id_{\State}$ which is the smallest element in the equivalence class of $f$
w.r.t. pointwise ordering, which is such that the
prefix points of $f$ are the roots of $f'$.
Viceversa by taking $f'' \DefEq f \cup \Id_{\State}$ we would have
that the set $\Prefix{f}$ coincides with the fix points of $f''$,
a remark which is at the basis of the proof in the subsequent
lemma. 
\fi

}\end{remark}

\begin{proposition}[Cofinality of Realizers Prefix Points]\label{prop:cofin}
If $r:\State\arrow \State$ is a realizer, then for all
$s\in\State$ there is $s'\in \Prefix{r}$ such that $s\sleq s'$,
namely $\Prefix{r}$ is cofinal in $\State$ (in particular it is
non empty).
\end{proposition}

\begin{proof} Given $s\in \State$ define the mapping $\sigma:\Nat\arrow \State$ by $\sigma(0) := s$
and $\sigma(i+1):= \sigma(i) \sqcup r(\sigma(i))$, which exists
because of the compatibility of $r$ with its argument. By construction $\sigma$ is a w.i.
sequence, hence by the convergence of $r$, $r\circ\sigma$ has a
limit $\sigma(i_0)$ for some $i_0$, that is $r(\sigma(i_0))=
r(\sigma(i_0+1))$. Then
\[\sigma(i_0+1) = \sigma(i_0) \join r(\sigma(i_0)) = \sigma(i_0) \join r(\sigma(i_0+1)),\]
which implies $r(\sigma(i_0+1)) \sleq \sigma(i_0+1) $; clearly $s
= \sigma(0)\sleq \sigma(i_0+1)\in\Prefix{r}$.
\end{proof}

\begin{remark}\label{rem:r-search}{\em
The proof of Proposition \ref{prop:cofin} describes a computation
that, given an arbitrary $s_0$, produces the w.i. sequence $\sigma(0) =
s_0, \sigma(1) = r(s_0) \join s_0, \sigma(2) = r(r(s_0) \join s_0)
\join r(s_0) \join s_0, \ldots$ until a prefix point $\sigma(n)$ is found.
Each time the sequence strictly increases, it is because
$r(\sigma(i)) \neq\bot$, that intuitively means that the realizer
$r$ has something to add to $\sigma(i)$ to reach its own goal,
abstractly constituted by the set of prefix points of $r$. This search
procedure, which is recursive in $r$, is monotonic as the
knowledge grows, but this happens because only positive
information is stored in the state. As a matter of fact the growth
of the sequence generated by $r$ might redefine the values of some
$\chiP$ and $\phiP$ occurring in a formula $A$, which is the
actual goal of $r$ when it is a realizer of $A$: this is an
implicit backtracking (more precisely $1$-backtracking: see \cite{BerCoqHay}),
in the sense that we are retracting
previous definitions of these symbols, and in particular of the
Skolem functions $\phiP$, until $A$ becomes true.

After some $i$ has been found such that $r(\sigma(i)) = \bot$, the whole construction stops.
However nothing prevents that, later, new atoms might be added to $\sigma(i)$, producing some 
$s'\sqsupseteq\sigma(i)$ not in $\Prefix{r}$. Now the cofinality
of $\Prefix{r}$ in $\State$ implies that we can resume
the search at $s'$ and that it will eventually
succeed in finding some other $s''\sqsupseteq s'$ which is in $\Prefix{r}$.}\end{remark}

The reaching of a goal is represented by finding a prefix point of the relative realizer. The next proposition says that the prefix points of a merge are exactly the prefix points common to both the merged realizers.

\begin{proposition}\label{prop:mergeOfRealizers}
Suppose that $\monoid$ is a merge: then for any pair of realizers
$r,r'$, $r\Smonoid r'$ is a realizer. Moreover:
\[\Prefix{r\Smonoid r'} = \Prefix{r} \cap \Prefix{r'}.\]
\end{proposition}

\begin{proof}
In view of the definition $\Smonoid =
\SMonad(\monoid)\circ\psi_{\State,\State}$, we know that
$r\Smonoid r'$ is strongly convergent by Corollary
\ref{cor:psi-convergent} and Remark \ref{rem:Sf-continuous}, since
$r$ and $r'$ are such.

For any $s\in\State$ we know that $\Compatible{s}{r(s)}$ and
$\Compatible{s}{r'(s)}$; now $(r\Smonoid r')(s) = r(s)\monoid
r'(s)$ for all $s\in\State$ and we conclude that
$\Compatible{s}{(r(s)\monoid r'(s))}$ by (\ref{lem:mergeMprop-a})
of Lemma \ref{lem:mergeMprop}.

By (\ref{def:realizer-a}) of Definition \ref{def:realizer},
$r(s)\cap s = \bot = r'(s)\cap s$; by (\ref{lem:mergeMprop-b}) of Lemma \ref{lem:mergeMprop}
this implies:
\[(r\Smonoid r')(s) \cap s = (r(s) \monoid r'(s))\cap s = \bot.\]
This concludes the proof that $r\Smonoid r'$ is a realizer.

This last fact implies that $\Prefix{r\Smonoid r'} =
\Set{s\in\State\mid r(s)\monoid r'(s) = \bot}$: by
(\ref{def:mergeM-b}) of Definition \ref{def:mergeM} we know that
$r(s)\monoid r'(s) = \bot$ implies both $r(s) = \bot$ and $r'(s) =
\bot$, namely that $\Prefix{r\Smonoid r'} \subseteq \Prefix{r}
\cap \Prefix{r'}$.

Viceversa, if $s\in \Prefix{r} \cap \Prefix{r'}$ then $r(s) = \bot
= r'(s)$, so that, by $\bot\monoid\bot = \bot$, we have that $r(s)
\monoid r'(s) = \bot$, that is $s\in\Prefix{r\Smonoid r'}$.
\end{proof}

We now relate formally interactive realizers to formulas of $\LangOne$.
First we define an abstract relation between realizers and families of sets 
indexed over $\State$ which we call {\em interactive forcing}.

\begin{definition}[Interactive Forcing]\label{def:forcing}
Let $r$ be a realizer, $\alpha \in\SMonad X$ and
$Y=\Set{Y_s \mid s\in\States}$ a family of subsets of $X$ indexed over $\State$. 
Then {\em
$r$ interactively forces $\alpha$ into $Y$}, written
$\forces{r}{\alpha}{Y}$, if for all $s\in\Prefix{r}$ it is the case that $\alpha(s)
\in Y_s$. 
\end{definition}

Let us now consider the formulas. 
In the standard model the semantics of a formula $A$ with (free)
variables included into $\vec{x} = x_1,\ldots, x_k$ is a $k$-ary
relation over $\Nat$, which is the extension of the formula. In our model the
{\em extension} of $A$ is the $\State$-indexed family of sets 
$\ext{A} \DefEq \Set{\ext{A}_s\mid s\in\States}$ where:
\[\ext{A}_s \DefEq \Set{\vec{m} \mid \fv{A} \subseteq \vec{x} \And |\vec{x}| = |\vec{m}| \And
\Sem{A}^{\SMonad }_{[\overrightarrow{\lambda\_.m}/\vec{x}]}(s)=\TrueVal}.\]
Here $\vec{m} = m_1,\ldots,m_k$ is a $k$-ple of natural numbers,
$|\vec{m}| = |\vec{x}| = k$, and $[\overrightarrow{\lambda\_.m}/\vec{x}] =
[\lambda\_\,.m_1/x_1,\ldots,$ $\lambda\_\,.m_k/x_k]$ is the
environment associating $\lambda\_\,.m_i$ to $x_i$ for each $i$. 
We now define the {\em forcing} of $A$ in terms of the extension of $A$.

\begin{definition}[Interactive Forcing of a Formula]\label{def:formula_forcing}
Let $r$ be a realizer, $A\in\LangOne$ with $\fv{A} \subseteq \vec{x} = 
x_1,\ldots, x_k$, and $\vec{\alpha} = \alpha_1,\ldots,\alpha_k \in\SMonad\Nat$.
Then we say that {\em $r$ interactively forces $\vec{\alpha}$ into $A$}, 
written $\forces{r}{\vec{\alpha}}{A(\vec{x})}$, 
if $\forces{r}{\tuple{\alpha_1,\ldots,\alpha_k}}{\ext{A}}$.
\end{definition}

To each formula $A\in\LangOne$ Definition \ref{def:formula_forcing} associates the
relation $\Set{(r,\vec{\alpha}) \mid \forces{r}{\vec{\alpha}}{A(\vec{x})}}\subseteq
\SMonad(\State) \times  \SMonad(\Nat^k) \simeq \RMonad(\Nat^k)$, where $k$
is the length of $\vec{\alpha}$ and $\vec{x}$, making apparent the connection between
forcing and the realizer monad $\RMonad$.


\medskip
In view of Proposition \ref{prop:mergeOfRealizers} and of Remark
\ref{rem:r-search}, the intuitive idea of the forcing relation $\forces{r}{\vec{\alpha}}{A(\vec{x})}$
is that, whenever the variables $\vec{x}$
including all the free variables of $A$ are interpreted by the
individuals $\vec{\alpha}$,  the
sequence generated by $r$ out of an arbitrary $s_0$ will eventually reach (in a
finite number of steps) some state $s\in\Prefix{r}$ making true 
that $\vec{\alpha}(s) \in \ext{A}_s$.
This is however a subtly complex task:
the action of $r$ is to direct $\vec{\alpha}$ into $\ext{A}$ by
extending the given state; but we must keep in mind that such a
search aiming at the target $\ext{A}_s$ for some $s$, moves the
target itself as a side effect. 
Note also that:
\[\tuple{\alpha_1,\ldots,\alpha_k}(s) =
(\alpha_1(s),\ldots,\alpha_k(s)) \in \ext{A}_s \Iff
\Sem{A}^{\SMonad
}_{[\overrightarrow{\lambda\_.\alpha(s)}/\vec{x}]}(s) =
\Sem{A}^{\SMonad }_{[\vec{\alpha}/\vec{x}]}(s) = \TrueVal.
\]
By the fact that we do not ask that the free variables of $A$ are
exactly $\vec{x}$, but only included among them, the sets
$\ext{A}_s$ contain tuples of different length (thought there is a
minimum length which is the cardinality of $\fv{A}$), which
implies that if $\forces{r}{\vec{\alpha}}{A(\vec{x})}$ then
$\forces{r}{\vec{\alpha},\vec{\beta}}{A(\vec{x},\vec{y})}$ for all
vectors $\vec{y}$ and $\vec{\beta}$ such that $|\vec{y}| = |\vec{\beta}|$.

\medskip
Toward the proof of the claim that any theorem of $\PRA+\EM$ is interactively realizable,
we begin with the logical and arithmetic axioms. Together we consider also the $(\varphi)$-axioms,
since in all these cases the realizer turns out to be the trivial one.

\begin{lemma}[Logical, Arithmetic or $(\varphi)$-Axioms]\label{lem:PRA+phi-axioms}
If $A$ is either a non logical axiom of $\PRA$, or an axiom of
$\IPC$, or an instance of the $(\varphi)$-axiom, then
$\forces{\lambda\_\,.\bot}{\vec{\alpha}}{A}$.
\end{lemma}

\begin{proof} It follows by Proposition
\ref{prop:conservatitvity}, since $\lambda\_\,.\bot$ is a realizer
and $\Prefix{\lambda\_\,.\bot} = \State$.
\end{proof}

Now we come to the study of the $(\chi)$-axioms.
For any $k+1$-ary primitive recursive predicate $\Pred{P}$ (we
abuse notation below, writing ambiguously $\Pred{P}$ for the
symbol and for its standard interpretation) let us define
$r_{\Pred{P}}:\Nat^{k+1}\times \State\arrow \State$ as follows:
\[r_{\Pred{P}}(\vec{m},n, s) = \left\{\begin{array}{ll}
    \Set{\tuple{\Pred{P},\vec{m},n}} & \mbox{if $\Pred{P}(\vec{m},n)$ and
        $\forall n'. \tuple{\Pred{P},\vec{m},n'} \not\in s$} \vspace{1mm} \\
    \bot & \mbox{else.}
\end{array}\right.\]

\begin{lemma}\label{lem:r_P}
For all $\vec{m}, n \in \Nat$, $\lambda s\in\State. \
r_{\Pred{P}}(\vec{m},n, s)$ is a realizer.
\end{lemma}

\begin{proof} That $r_{\Pred{P}}(\vec{m},n, s)\cap s = \bot$ for any $s\in\State$ is
immediate by definition. It remains to prove that $\lambda s\in\State.\;r_{\Pred{P}}(\vec{m},n,s)$ is strongly
convergent and consistent with its arguments.

Let $\sigma$ be any w.i. sequence. If $\neg \,\Pred{P}(\vec{m},n)$ then
$r_{\Pred{P}}(\vec{m},n, \sigma(i)) = \bot$ for all $i$. Suppose
instead that $\Pred{P}(\vec{m},n)$ is true. If
$\tuple{\Pred{P},\vec{m},n'}\not\in \sigma(i)$ for all $n'$ and
$i$, then $r_{\Pred{P}}(\vec{m},n, \sigma(i)) =
\Set{\tuple{\Pred{P},\vec{m},n}}$ for all $i$; otherwise there
exist $i$ and $n'$ such that for all $j\geq i$,
$\tuple{\Pred{P},\vec{m},n'}\in\sigma(j)$, as $\sigma$ is weakly
increasing. Then $r_{\Pred{P}}(\vec{m},n, \sigma(j)) = \bot$ for
all $j\geq i$.

If $r_{\Pred{P}}(\vec{m},n, s) = \Set{\tuple{\Pred{P},\vec{m},n}}$
then $\Pred{P}(\vec{m},n)$ is true so that
$\Set{\tuple{\Pred{P},\vec{m},n}}\in \States$. Moreover
$\tuple{\Pred{P},\vec{m},n'} \not\in s$ for all $n'\in\Nat$, hence
$\Compatible{\Set{\tuple{\Pred{P},\vec{m},n}}}{s}$. If instead
$r_{\Pred{P}}(\vec{m},n, s) = \bot$ then the thesis holds
trivially since $\Compatible{\bot}{s}$.
\end{proof}

Recall that $r_{\Pred{P}}^{\SMonad} = \Sext{r_{\Pred{P}}} \circ
\psi_{\Nat^{k+1}}$, so that for any
$\vec{\alpha},\beta\in\SMonad\Nat$ we have:
\[r_{\Pred{P}}^{\SMonad}(\vec{\alpha}, \beta) = \lambda s \in \State. r_{\Pred{P}}(\vec{\alpha}(s),\beta(s),s).\]
The next lemma states that, under the condition that
$\vec{\alpha},\beta$ are individuals (that is strongly
convergent), $r_{\Pred{P}}^{\SMonad}(\vec{\alpha}, \beta)$ is a
realizer of the $(\chi)$-axiom instance relative to $\Pred{P}$.

\begin{lemma}[$\chi$-Axiom]\label{lem:chi-axiom}
If $\Pred{P}$ is a $k+1$-ary primitive recursive predicate, and
$\vec{\alpha},\beta\in\SMonad\Nat$ are strongly convergent then
$r_{\Pred{P}}^{\SMonad}(\vec{\alpha}, \beta)$ is a realizer, and
it is such that:
\[\forces{r_{\Pred{P}}^{\SMonad}(\vec{\alpha}, \beta)}{\vec{\alpha}, \beta}{\Pred{P}(\vec{x},y)\Then \chiP(\vec{x})}.\]
\end{lemma}

\begin{proof} By definition, $r_{\Pred{P}}^{\SMonad}(\vec{\alpha},
\beta,s)= r_{\Pred{P}}(\vec{\alpha}(s),\beta(s),s)$, so that it is
consistent with $s$ and the intersection
$r_{\Pred{P}}^{\SMonad}(\vec{\alpha}, \beta,s)\cap s = \bot$
because $r_{\Pred{P}}(\vec{m},n, s)\cap s = \bot$ for all
$\vec{m},n$. Since $\Sext{r_{\Pred{P}}}$ is global by Corollary
\ref{cor:ext-pointwise}, we know that $r_{\Pred{P}}^{\SMonad}$ is
$k+1$-global by Lemma \ref{lem:k-global}. Now
\[r_{\Pred{P}}^{\SMonad}(\overrightarrow{\lambda\_\,.m},\lambda\_\,.n,s) =
r_{\Pred{P}}(\vec{m},n, s),\] and the latter is an individual by
Lemma \ref{lem:r_P}. It follows that
$r_{\Pred{P}}^{\SMonad}(\vec{\alpha}, \beta)$ is an individual if
$\vec{\alpha}$ and $\beta$ are such, by Corollary
\ref{cor:k-global}. We conclude that
$r_{\Pred{P}}^{\SMonad}(\vec{\alpha}, \beta)$ is a realizer.

If $s\in \Prefix{r_{\Pred{P}}^{\SMonad}(\vec{\alpha}, \beta)}$
then $r_{\Pred{P}}(\vec{\alpha}(s),\beta(s),s) = \bot$. It follows
that either $\neg \Pred{P}(\vec{\alpha}(s),\beta(s))$ or
$\tuple{\Pred{P},\vec{\alpha}(s),n} \in s$ for some $n\in\Nat$
(not necessarily equal to $\beta(s)$): this implies that
$\Sem{\chiP}^{\SMonad}_{[\vec{\alpha},\beta/\vec{x},y]}(s) =
\Sem{\chiP}(\vec{\alpha}(s),s) = \TrueVal$. In both cases we have
$\Sem{\Pred{P}(\vec{x},y)\Then
\chiP(\vec{x})}^{\SMonad}_{[\vec{\alpha},\beta/\vec{x},y]}(s) =
\TrueVal$, that is $\vec{\alpha}(s),\beta(s) \in
\ext{\Pred{P}(\vec{x},y)\Then \chiP(\vec{x})}_s$.

\end{proof}

\begin{remark}\label{rem:interaction}{\em
By reading Lemma \ref{lem:chi-axiom} together with Remark
\ref{rem:r-search} we see one reason for naming ``interactive
realizer'' the map $r = r_{\Pred{P}}^{\SMonad}(\vec{\alpha},
\beta)$. In fact we have seen that each time the sequence
generated by $r$ strictly increases it is because $r(\sigma(i))
\neq\bot$, and this happens whenever
$\Pred{P}(\vec{\alpha}(\sigma(i)),\beta(\sigma(i)))$ holds but
$\tuple{\Pred{P},\vec{\alpha}(\sigma(i)),n} \not\in \sigma(i)$ for
any $n\in\Nat$. In such a case the next state in the sequence is
$\sigma(i+1) = r(\sigma(i)) \join \sigma(i) =
\Set{\tuple{\Pred{P},\vec{\alpha}(\sigma(i)),\beta(\sigma(i))}}
\cup \sigma(i)$ so that the newly found tuple
$\tuple{\Pred{P},\vec{\alpha}(\sigma(i)),\beta(\sigma(i))}$ is
added to $\sigma(i)$. In particular if a prefix point of $r$ is
reached, that is $r(s) = \bot$, no more information is needed to
make the realized formula true w.r.t. such a state.

If we further consider the implication $\Pred{P}(\vec{x},y) \Then
\Pred{P}(\vec{x},\phiP(\vec{x}))$, which follows in $\PRA+\EM$ by
the $(\varphi)$-axiom and the $(\chi)$-axiom, we see that whenever
\[\Sem{\Pred{P}(\vec{x},y) \Then
\Pred{P}(\vec{x},\phiP(\vec{x}))}^{\SMonad}_{[\vec{\alpha},\beta/\vec{x},y]}(s)
= \FalseVal,\] the tuple
$\tuple{\Pred{P},\vec{\alpha}(s),\beta(s)}$ is the witness of the
fact that the implication fails at $s = \sigma(i)$. This is used
at the next step in the attempt to make it true at $s' = r(s)\join
s$, by means of the fact that
$\Sem{\phiP(\vec{x})}^{\SMonad}_{[\vec{\alpha}/\vec{x}]}(s') =
\beta(s)$. By this we see how the counterexample to the
implication at a previous stage is used to redefine the value of
the Skolem function $\phiP$ in the point $\vec{\alpha}(s')$, which
is the result of the interaction between the realizer $r$ and the
``nature'', that is the standard model.

However it is not necessarily the case that $\vec{\alpha}(s') =
\vec{\alpha}(s)$, which implies that the value of
$\Sem{\Pred{P}(\vec{x},y) \Then
\Pred{P}(\vec{x},\phiP(\vec{x}))}^{\SMonad}_{[\vec{\alpha},\beta/\vec{x},y]}(s')$
could still be $\FalseVal$. It is here that the hypothesis that
the $\vec{\alpha}$ are convergent is crucial, since $s\sleq s'$
and in general the value of $\vec{\alpha}\circ \sigma$ is
eventually constant. }\end{remark}

According to our interpretation, logical rules are realized by the merging of the realizers of the premises.
Let us first consider the case of modus ponens.

\begin{lemma}[Modus Ponens Rule]\label{lem:modusPonens}
If $\forces{r}{\vec{\alpha}}{A}$ and
$\forces{r'}{\vec{\alpha}}{A\Then B}$ then $\forces{r \Smonoid
r'}{\vec{\alpha}}{B}$.
\end{lemma}
\begin{proof} Let $\vec{\alpha} = \alpha_1,\ldots,\alpha_k$: then
$\ext{A\Then B}_s,\ext{A}_s$ and $\ext{B}_s$ are subsets of the
universe $\Nat^k$, so that in particular we can take the
complement $\overline{\ext{A}_s} = \Nat^k\setminus \ext{A}_s$.
Then let us observe that for all $s\in\State$:
\[ \ext{A\Then B}_s = \overline{\ext{A}_s} \union \ext{B}_s.\]
By Proposition \ref{prop:mergeOfRealizers} we know that $r
\Smonoid r'$ is a realizer such that $\Prefix{r \Smonoid r'} =
\Prefix{r}\cap\Prefix{r'}$. Therefore, by the hypotheses, if
$s\in\Prefix{r \Smonoid r'}$ then
\[ \vec{\alpha}(s) \in \ext{A}_s \intersection \ext{A\Then B}_s  =
  \ext{A}_s \intersection (\overline{\ext{A}_s} \union \ext{B}_s) =
  \ext{A}_s \intersection \ext{B}_s, \]
hence $\vec{\alpha}(s) \in \ext{B}_s$ as desired.
\end{proof}

\begin{remark}\label{rem:MP-merge}{\em In Remark
\ref{rem:interaction} we have stressed that even in the case of
the $(\chi)$-axiom the realizer might reach its prefix point in
several steps, and after several tests against the standard model
of arithmetic: this is a first form of interaction. The case of
modus ponens rule MP, as well as the more complex one of IND
treated below, illustrates  a second form of interaction between two or more realizers.
While searching a prefix point of $r \Smonoid r'$ the given realizers
$r$ and $r'$ do not necessarily move to the same states, not even
to compatible ones. The realizer $r \Smonoid r'$ let
$r$ and $r'$ to dialogue via the state by merging of the
respective sequences they generate. This
process depends on the choice of the merge: with
$\monoid^{\SMonad}_0$, for example, it is a rigid interleaving of
the searches generated by $r$ and $r'$, giving precedence to $r$,
while with $\monoid^{\SMonad}_1$ and $\monoid^{\SMonad}_2$ the
resulting sequence is generated by parallel
process.

We observe that the merging of realizers is the meaning of any inference
rule with more than one premise.}\end{remark}

Recall the convention that the writing $A(x)$ means that $x$ might
occur free in $A$, and $A(t)$ is informal for the substitution
$A[t/x]$ of $t$ for $x$ in $A$.

\begin{lemma}[Substitution Rule]\label{lem:subRule}
If $\forces{r}{\vec{\alpha},\beta}{A(\vec{x},y)}$ for all
convergent $\vec{\alpha},\beta$, then for any $t\in\LangOne$ such
that $\fv{t}\subseteq\vec{x}$,
$\forces{r}{\vec{\alpha},\beta}{A(\vec{x},t)}$.
\end{lemma}

\begin{proof}
By the hypothesis and the fact that
$\Sem{t}^{\SMonad}_{[\vec{\alpha}/\vec{x}]}$ is convergent by
Theorem \ref{thr:convInterp}, we have that
$\forces{r}{\vec{\alpha},\Sem{t}^{\SMonad}_{[\vec{\alpha}/\vec{x}]}}{A(\vec{x},y)}$,
where we note that the environment $[\vec{\alpha}/\vec{x}]$ is not
defined over $y$, which however does not occur in $t$. By Lemma
\ref{lem:substitution}
\[\Sem{A(\vec{x},y)}^{\SMonad}_{[\vec{\alpha},\Sem{t}^{\SMonad}_\xi/\vec{x},y]} =
  \Sem{A(\vec{x},t)}^{\SMonad}_{[\vec{\alpha}/\vec{x}]},\]
so that $\forces{r}{\vec{\alpha}}{A(\vec{x},t)}$ and, since
$y\not\in\fv{A(\vec{x},t)}$, also
$\forces{r}{\vec{\alpha},\beta}{A(\vec{x},t)}$.
\end{proof}

\begin{lemma}[Induction Rule]\label{lem:induction}
Suppose that for all convergent $\vec{\alpha}$ and $\beta$:
\[\forces{r(\vec{\alpha})}{\vec{\alpha}}{A(x,\Fun{0})}
~~~\mbox{and}~~~
\forces{r'(\vec{\alpha},\beta)}{\vec{\alpha},\beta}{A(x,y)\Then
A(\vec{x},\Succ(y))}.\] For all $\vec{\alpha}$ let
$f(\vec{\alpha}):\Nat\arrow\SMonad(\State)$ be defined by
(primitive) recursion: $f(\vec{\alpha},0) = \lambda \_\,.\bot$ and
$f(\vec{\alpha},n+1) = f(\vec{\alpha},n) \Smonoid
r'(\vec{\alpha},\lambda\_\,.n)$. Then for all convergent
$\vec{\alpha}$ and $\beta$, $\Sext{f(\vec{\alpha})}(\beta)$ is a
realizer and:
\[\forces{r \Smonoid
(\Sext{f(\vec{\alpha})}(\beta))}{\vec{\alpha},\beta}{A(x,y)}.\]
\end{lemma}

\begin{proof} To simplify the notation,
we fix the vector $\vec{\alpha}$ and write just $r$ for
$r(\vec{\alpha})$, $r'(\beta)$ for $r'(\vec{\alpha},\beta)$,
$f(n)$ for $f(\vec{\alpha},n)$ and hence $\Sext{f}(\beta)$ for
$\Sext{f(\vec{\alpha})}(\beta)$.

First we have to check that $\Sext{f}(\beta)$ is a realizer. Note
that for any $n\in\Nat$ we have $\Sext{f}(\lambda\_\,.n) =
r'(\lambda\_\,.0) \Smonoid \cdots \Smonoid r'(\lambda\_\,.n-1)$
(or just $\Lbot$ when $n = 0$), which is a realizer by Proposition
\ref{prop:mergeOfRealizers}. The function $\Sext{f}(\beta)$ is
global (or $k$-global to take the $\vec{\alpha}$ into account) by
Corollary \ref{cor:ext-pointwise} and, as we have just seen, it
sends constant individuals into realizers which are individuals of
$\State$: hence $\Sext{f}(\beta)$ is an individual for any
individual $\beta$ by (\ref{thr:density-b}) of Theorem
\ref{thr:density}. The remaining conditions (\ref{def:realizer-b})
and (\ref{def:realizer-c}) of Definition \ref{def:realizer} are
immediately seen to hold by observing that for all $s\in\State$,
$\Sext{f}(\beta,s) = r'(\lambda\_\,.0,s) \monoid \cdots \monoid
r'(\lambda\_\,.\beta(s)-1,s)$.

In order to prove the thesis, we establish by induction over $n$ that:
\begin{equation}\label{eq:induction}
\forall n \in \Nat.\; \forces{r \Smonoid
\Sext{f}(\lambda\_\,.n)}{\vec{\alpha},\lambda\_\,.n}{A(x,y)}.
\end{equation}

For the base case we have $r \Smonoid \Sext{f}(\lambda\_\,.0) = r
\Smonoid \Lbot = r$, and we know that
$\forces{r}{\vec{\alpha}}{A(x,\Fun{0})}$, which implies
$\forces{r}{\vec{\alpha},\lambda\_\,.0}{A(x,\Fun{0})}$ vacuously
as $y\not\in\fv{A}$.

For the step case we have $r \Smonoid \Sext{f}(\lambda\_\,.n+1) =
r \Smonoid \Sext{f}(\lambda\_\,.n) \Smonoid r'(\lambda\_\,.n)$,
but:
\[\begin{array}{ll}
\forces{r'(\lambda\_\,.n)}{\vec{\alpha},\lambda\_\,.n}{A(x,y)\Then
A(\vec{x},\Succ(y))} & \mbox{by the hypothesis of the lemma, and} \vspace{2mm} \\
\forces{r \Smonoid
\Sext{f}(\lambda\_\,.n)}{\vec{\alpha},\lambda\_\,.n}{A(x,y)} &
\mbox{by induction hypothesis.}
\end{array}\]
We then obtain that $\forces{r \Smonoid
\Sext{f}(\lambda\_\,.n+1)}{\vec{\alpha},\lambda\_\,.n}{A(x,\Succ(y))}$,
by Lemma \ref{lem:modusPonens}. By the Substitution Lemma
\ref{lem:substitution},
$\Sem{A(x,\Succ(y))}^{\SMonad}_{[\vec{\alpha},\lambda\_\,.n/\vec{x},y]}
=
\Sem{A(x,y)}^{\SMonad}_{[\vec{\alpha},\lambda\_\,.n+1/\vec{x},y]}$,
and therefore we conclude that $\forces{r \Smonoid
\Sext{f}(\lambda\_\,.n+1)}{\vec{\alpha},\lambda\_\,.n+1}{A(x,y)}$.

Now for any $\beta\in\SMonad\Nat$ and $s\in\State$:
\[(r \Smonoid \Sext{f}(\beta))(s) = r(s) \monoid \Sext{f}(\beta,s)
= r(s) \monoid \Sext{f}(\lambda\_\,.\beta(s),s),
\]
because $\Sext{f}$ is global, and $\forces{r \Smonoid
\Sext{f}(\lambda\_\,.\beta(s))}{\vec{\alpha},\lambda\_\,.\beta(s)}{A(x,y)}$
by (\ref{eq:induction}) above since $\beta(s)\in\Nat$. It follows
that if $s\in\Prefix{r \Smonoid \Sext{f}(\beta)}$ then
\[(r \Smonoid \Sext{f}(\beta))(s) = \bot = (r \Smonoid
\Sext{f}(\lambda\_\,.\beta(s)))(s),\] so that $s\in\Prefix{r
\Smonoid \Sext{f}(\lambda\_\,.\beta(s))}$. This implies that
\[ \Sem{A(x,y)}^{\SMonad}_{[\vec{\alpha},\beta/\vec{x},y]}(s) =
\Sem{A(x,y)}^{\SMonad}_{[\vec{\alpha},\lambda\_\,.\beta(s)/\vec{x},y]}(s) =
\TrueVal,
\]
by Lemma \ref{lem:k-globalEnv}.
\end{proof}

\begin{remark}\label{rem:IND-realizer}{\em
The point of the the proof of Lemma \ref{lem:induction} is the use
of Density Theorem \ref{thr:density}. In fact the interpretation
of induction via primitive recursion implies that we are proving
some statement about numbers in $\Nat$ and that we are able to
compute with them, while the realizability interpretation deals
with individuals in $\SMonad\Nat$. The import of density is that,
given that the interpretation of formulas is a global function of
the individuals interpreting their variables, everything lifts
uniformly from $\Nat$ to $\SMonad\Nat$.}\end{remark}

\begin{theorem}[Interactive Realizability Theorem]\label{thr:realizability}
Suppose that $\PRA+\EM\der A$, for some $A\in\LangOne$ with
$\fv{A}\subseteq \vec{x} = x_1,\ldots,x_k$. Then for all
$\vec{\alpha} = \alpha_1,\ldots,\alpha_k$ of individuals in
$\SMonad\Nat$ there exists a realizer $r(\vec{\alpha})$ which is
recursive in $\vec{\alpha}$, such that
$\forces{r(\vec{\alpha})}{\vec{\alpha}}{A}$. Moreover the form of
$r(\vec{\alpha})$ depends on the proof of $A$ in $\PRA+\EM$.
\end{theorem}

\begin{proof} The existence of $r(\vec{\alpha})$ follows by the
lemmas \ref{lem:PRA+phi-axioms}, \ref{lem:chi-axiom},
\ref{lem:modusPonens}, \ref{lem:subRule} and \ref{lem:induction},
and by the remark that (possibly after renaming) the length $k$ of
$\vec{x}$ and $\vec{\alpha}$ can be taken to be large enough to
include all variables occurring in the proof. That $r$ is a
recursive functional of $\vec{\alpha}$ follows by the fact that
all realizers constructed in the lemmas above are
$\lambda$-definable. Finally that $r(\vec{\alpha})$ (and hence $r$
itself) actually reflects the structure of the proof of $A$ is
clear by construction.

\end{proof}

By choosing as the convergent $\vec{\alpha}$ the vector of
constant individuals $\overrightarrow{\lambda\_\,.m}$ one can use
the realizer associated to the proof of a $\PRA+\EM$ theorem to
compute the witness $\phiP(\vec{t}\;)$ of $A(\phiP(\vec{t}\;))$ in
the standard input $\vec{m}$: the general case of convergent
$\vec{\alpha}$ is however needed in the proof of the theorem and
for the compositionality of its construction.

The computational content of the proof, which we identify with the
realizer, is however trivial in case no instance of the
$(\chi)$-axiom occurs in it.

\begin{corollary}\label{cor:PRA-phi}
If $\PRA+(\varphi)\der A$ for $A\in\LangOne$, then the realizer
$r(\vec{\alpha})$ of Theorem \ref{thr:realizability} is just
$\Lbot$. Hence $\Sem{A}^{\SMonad }_\xi(s) = \TrueVal$ for all
environment $\xi$ and state $s$.
\end{corollary}

\begin{proof}
By inspection of the lemmas used in the proof of the theorem, the
realizer $r(\vec{\alpha})$ is a composition of the realizers used
in the axioms by the operator $\Smonoid$: this is immediate in all
cases but for induction rule (Lemma \ref{lem:induction}), which is
however easily checked by induction over $\Nat$, and then lifted
to $\SMonad\Nat$ by the very same argument we used in the proof of
the lemma. But in case the proof of $A$ does not use any instance
of the $(\chi)$-axiom, they are all $\Lbot$, which is the unit of
$\Smonoid$.

Now the second part of the thesis follows by Theorem
\ref{thr:realizability}, and by remembering that $\Prefix{\Lbot} =
\State$.
\end{proof}

\section{Related works}
\label{sec:related}
The spectrum of research ideas that have been influential on our work and of those that, even a posteriori, reveal to be connected to our results is too wide to be exhaustively treated. Therefore we limit ourself to a sketch of the approaches we feel closer to ours, either because we are building over them, or because we want to underline similarities and differences.

\medskip
{\em Coquand's game semantics of classical arithmetic.}
A primary source of the present research is Coquand's semantics of evidence for classical arithmetic \cite{Coq91}. 
Non constructive principles like excluded middle are treated there by means of backtracking and learning, and rely on the fact
that in each play only a finite amount of information about them is actually needed. The concept of 1-backtracking games appearing in \cite{BerCoqHay}, which can be seen as a restricted form of games in  \cite{Coq91} but with plays of possibly infinite length, is closely related to the present work. Given a game $G$, the 1-backtracking game $\mbox{\tt bck}(G)$ allows the player to come back to some previous move in a play, by undoing and forgetting all the intermediate moves made by either players. In this modified game the player does not loose if a loosing position is reached, rather the player looses only if forced to backtrack to the same position infinitely many times. This backtracking procedure, in which a player's strategy over  $\mbox{\tt bck}(G)$ consists, can be interpreted as a learning procedure, in the sense that the player learns from her own trials and errors, and is allowed to follow a different path in a tree of plays using the experience made so far. A winning strategy over $\mbox{\tt bck}(G)$ is effective, and the truth of any $\PRA+\EM$ theorem can be learned in this way: this is not true, however, if non constructive principles are admitted of logical complexity which is higher than $\EM$.

We might interpret our construction as an implementation of the same idea. This has been explained in the text, and especially in the remarks of Section \ref{sec:realizTheorem}. With respect to  \cite{BerCoqHay} we provide the needed machinery, together with a language, to denote learning strategies which is tailored for extracting them out of proofs.

\medskip
{\em Gold's theory of learning in the limit.}
The concept of dynamic individuals comes from Gold's theory of learning in the limit, exposed
in \cite{Gold65,Gold67}. A $k$-ary numerical function $f$ is {\em computable in the limit} if there exists a recursive (total)  $k+1$-function $g$ such that for all $\vec{m}\in\Nat^k$ the sequence 
$g(\vec{m},0), g(\vec{m},1),\ldots$ is eventually constant and equal to $f(\vec{m})$, i.e. $f(\vec{m}) = \lim_{n\in\Nat} g(\vec{m},n)$.
Then $g$ is a {\em guessing function} for $f$, that makes the values of $f$ learnable. This is equivalent to say that $f(\vec{m}) $
is an individual in our sense. Viceversa
the functions  $\phiP(\vec{m})$ and the predicates $\chiP(\vec{m})$ are computable in the limit
(with the $(k+1)$-th argument in $\State$, but note that states are concrete and finite objects, hence encodable into $\Nat$) by their guessing functions $\Sem{\phiP}$ and $\Sem{\chiP}$ (see Definition \ref{def:chi-phi-interp} above), with the minor difference that we take the limit w.r.t. w.i. sequences over
$\State$ (but note that states are concrete and finite objects and that $\State$ is a decidable set, hence encodable into a decidable subset of $\Nat$). More importantly, the ordering
of $\State$ is state extension rather than the arbitrary ordering of (code) numbers. Moreover infinitely many incompatible w.i. sequences  exist in $\State$, hence the limit of an individual depends on the choice of the  sequence, in general.

\medskip
{\em The $\varepsilon$-substitution method.} This is a method to eliminate quantifiers, by replacing them with 
$\varepsilon$-terms, which has been introduced in \cite{HilbBern70} (but see the exposition and improvement of the method in \cite{Mints96}). It consists in the introduction of the new term
$\varepsilon x A$ for each formula $A(x)$, whose meaning is: ``the least $x$ such that $A(x)$'', that makes quantifiers definable by adding the new axioms (called {\em critical formulas}): $A(t)\rightarrow A(\varepsilon x A)$. Any classical proof of arithmetic can be transformed into a proof without quantifiers and using as axioms a finite set of critical formulas instead; now Hilbert suggested and Ackermann proved that one can effectively find a {\em solving substitution} $S$ of $\varepsilon$-terms by numerals validating all the formulas in the proof (for which it suffices to validate the critical formulas). This is achieved by arranging a sequence $S_0,S_1,\ldots$ of substitutions such that $S_0$ is the identically $0$ substitution, and $S_{i+1}$ is obtained from $S_i$ as follows: choose an axiom $A(t)\rightarrow A(\varepsilon x A)$ in the proof such that $S_i(A(t)) = {\sf true}$ and $S_i(A(\varepsilon x A)) = {\sf false}$ if any (in the negative case $S = S_i$ and we are done).
Then put $S_{i+1}(\varepsilon x A) = \mbox{the least $n\leq S_i(t)$ such that}\; A(n)$. 

This is strikingly similar to the action of the realizers of the $(\chi)$-axioms, for which we refer to Remark \ref{rem:interaction}.
But this also reveals a key difference with our construction: in fact to avoid circularities, after redefining the value of
$\varepsilon x A$ one has to reset to $0$ the values of all the $\varepsilon$-terms of greater rank than 
$\varepsilon x A$ (a measure of the nesting of 
$\varepsilon$-terms). This indiscriminate form of backtracking, which is not very different from blind search, is the consequence of the limited use of information from the the proof in the $\varepsilon$-substitution method, that contributes only to determine the set of critical formulas that have to be satisfied. On the contrary the compositional nature of our realizability interpretation allows for an essential use of the proof structure, so that the nature and efficiency of the resulting algorithm strictly depends on the the proof itself, often embodying clever computational ideas.

\medskip
{\em Friedman $A$-translation.} Friedman's famous result in \cite{Fri78} is an extension of G\"odel proof of conservativity of $\PA$ over $\HA$ for $\Pi^0_2$-statements, which is suitable for program extraction from classical proofs. As such, it has been developed into a system to synthesise programs from classically provable $\Pi^0_2$-statements, or equivalently $\Sigma^0_1$-formulas possibly with parameters: see e.g. \cite{BergerBS02} and the MINLOG project. The extraction process runs as follows: given a classical proof $p$ of the arithmetic formula $A = \exists y\; \Pred{P}(x,y)$, with $\Pred{P}$ quantifier free (or equivalently primitive recursive) it is translated into a proof $p'$ in minimal arithmetic $\MA$ (i.e. $\HA$ without the axiom schema $\bot\rightarrow B$), of the formula $A^{\neg\neg}$, which is obtained from $A$  by double negation of atomic subformulas and interpreting $\Or,\exists$ by $\neg\wedge\neg$ and $\neg\forall\neg$ respectively. The translation is possible because, under such interpretation, the excluded middle law is trivially derivable, and in fact it is an instance of the identity rule.
Since $\neg B = B \rightarrow \bot$, we have $A^{\neg\neg} = \forall y (\Pred{P}(x,y) \rightarrow \bot) \rightarrow\bot$.
On the other hand by the absence of the ex-falso quodlibet law from $\MA$, $\bot$ can be replaced by an arbitrary formula, so that in particular we have a proof $p''= p'[A/\bot]$ of the formula $A[A/\bot] = \forall y (\Pred{P}(x,y) \rightarrow A) \rightarrow A$. Now $p''$ is a constructive proof, and $A[A/\bot]$ is constructively equivalent to $A$: hence we extract a program (a $\lambda$-term) from $p''$ realizing $\exists y\; \Pred{P}(x,y)$ which, for any $x$, actually computes a $y$ s.t. $\Pred{P}(x,y)$.

Apparently Friedman's interpretation bears no relation with our construction. However, if we consider only proofs of $A$ in $\EM$ and analyse the reduction paths from $p''$ we see that it computes the witness in a way conceptually similar to ours. For any term $t$ if $p''$ includes a subproof $q = \lambda\xi.q'[t/x]$ of  $\forall y (\Pred{P}(t,y) \rightarrow A)$, then $p''$ corresponds to the assumption that $\forall y \neg \Pred{P}(t,y)$ holds, and essentially stores the current state of the program. The term $q$ is what is called a ``continuation'' in functional programming. In fact, if for some $n$ a proof $r$ of
$\Pred{P}(t,n)$ is found, then $p''$ yields $q(r)$, that reduces to $q'[t/x,r/\xi]$, which actually restarts the computation from the point in which the wrong assumption $\forall y \neg \Pred{P}(t,y)$ was made, and, at the same time, produces a proof of $\exists y.\,\Pred{P}(t,y)$ and a witness $n$. According to our construction the same effect is achieved by adding $\Pred{P}(t,n)$ to the state. 

We observe that, beside being semantically more perspicuous, the method proposed in the present paper is even more general: the algorithm obtained via the functional interpretation of the Friedman's translation is essentially sequential and deterministic, the latter being an accidental and arbitrary feature; on the other hand with the present construction realizers, and in particular the merge operation that they embody, can be implemented in many different ways, also including parallel and non deterministic procedures.

\medskip
{\em Realizability of classical logic and theories.}
The extension of Kleene's realizability and of the Curry-Howard correspondence between proofs and programs to classical logic and theories began with Griffin's discovery, illustrated in \cite{Gri90}, that control operators and continuations can be typed by the classical law: $\neg\neg A \rightarrow A$. Since then this idea has been pursued by several authors: see e.g. \cite{Mur91} dealing however with Freedman's translation, and \cite{KrivineRealizability}, which is directly inspired to the Curry-Howard correspondence. A connected development has been by introducing calculi that are to classical proofs what the $\lambda$-calculus is for intuitionistic ones: see Parigot's $\lambda\mu$-calculus in \cite{Parigot92}, and its development into a symmetric calculus of classical proofs called $\lambda\mu\tilde{\mu}$-calculus in \cite{CurHerb-07}, but also the former ``symmetric $\lambda$-calculus'' in \cite{BarbaneraB96},  exploiting similar ideas.
With respect to all such works we have departed here because of the concept of realizer we have used, which is not based on the idea of continuations nor on that of control operators, but on searching and learning. However, and a posteriori, we think that similar remarks could be done as in the case of Friedman translation, namely that, on the one hand, we might expect to obtain similar algorithms on particular examples; but on the other we have a better explanation of how and why the interpretation works, even if, to the time, the present results are limited to proofs of low logical complexity, as repeatedly explained in the previous sections, while the above mentioned systems deal with full classical logic and arithmetic.

In \cite{Hay06} a notion of realizability is introduced for a subclassical arithmetic, called Limit Computable Arithmetic (LCM). 
The theory combines Kleene realizabiliy with Gold learning in the limit, which is achieved by 
asking a realizer to learn the evidence of the realized formula, instead of computing such an evidence.
The essential departure from Kleene's realizability is of course in the cases of $A \vee B$ and of $\exists x.\, A(x)$: according to Hayashi a realizer of $A\vee B$ is pair $g(n) = (g_1(n),g_2(n))$ such that $\lim_n g(n)$ exists and if $\lim_n g_1(n) = 0$ then $g_2$ is a guessing function for $A$, while it is a guessing function of $B$ in case $\lim_n g_1(n) \neq 0$. Similarly a realizer of
$\exists x.\, A(x)$ is  a function $g(n) = (g_1(n),g_2(n))$ that always converges and $g_2$ is guessing function for
$A(\lim_n g_1(n))$.

Hayashi's concept of realizer is similar to ours under relevant respects: it is constructed along the proof (and by this it is called the proof ``animator''), and it is convergent. The fact that LCM uses quantified formulas while $\PRA+\EM$ is a quantifier free theory is a minor difference, and not a true limitation of our approach: see  \cite{AschieriB09}. Rather the essential difference among
the learning realizability of LCM and the model we present here lies in the use of the proof, and hence of the realizer itself. In the case of LCM a realizer is a guessing function, hence a tool for testing guesses which have to be provided by a ``user'' interacting with the proof; in the absence of the user the only strategy to learn the truth of the conclusion of the proof is exhaustive blind search. On the contrary an interactive realizer in our sense is the basic block of a learning strategy, capable to produce
and test hypothesis against the ``nature'', namely that part of the standard model that can be learned within a finite number of steps.

In \cite{AschieriB09} essentially the same model that we have studied here is combined with Kleene's realizability obtaining
an interactive realizability interpretation of $\HA+\EM$. This extension of the interactive realizability model makes the terming ``realizability'' even more acceptable for the construction we are proposing.

However with respect to that work, we take here a different research direction: 
first we isolate and investigate on their own the concepts of individuals, global functions, interactive realizers and merge of realizers. These concepts, that are at the hearth of the construction, are somewhat hidden in the presence of nested quantifiers, 
that for example enforce the interpretation of a formula to be of different type depending on the formula itself; as a matter of fact 
in \cite{AschieriB09} the type of $\Sem{A}$ is $\State\arrow\Bool$ only in the case if atomic formulas; consequently also the type of realizers gets arbitrarily complex. More, we think that the framing of our model in the theory of strong monads, which is a major contribution of the present paper, allows a more general view of the construction and hints to its possible extensions to cope with non constructive principles of higher complexity.

\medskip
{\em Monads and the interpretation non-constructive proofs.}
Monads come from category theory, and strong monads have been introduced into the world of
typed $\lambda$-calculi and of the foundation of programming languages 
in \cite{Mog91}, where the reader will find the definitions of side-effects and 
continuations monads, but not of  the monads we use here: for that reason we described 
the monad $\SMonad$ in some detail, even if it is a quite simple example
of  strong monad.

As it should be clear from the text, we do not make essential use of
categorical techniques in our work, and base the exposition on
simply typed $\lambda$-calculus. This is coherent with Moggi's
original presentation of monads as type constructors of a
computational $\lambda$-calculus, and with the similar treatment
of this topic e.g. in the book \cite{AC:domlc}, chapter 8.

Coquand pointed out in
\cite{Coquand96computationalcontent} 
a suggestive connection between the constructive interpretation of classical principles
and monads, in which monads play for non
constructive features the same role that they have for simulating
imperative aspects into functional programming languages. 
 Indeed the monad $\SMonad$ is here the
main tool for defining formulas interpretation in a non ad-hoc
fashion, providing a nice characterization of global functions in
terms of morphisms of the Kleisli category $\CatSet_{\SMonad}$. It
is while attempting to devise the right definition of the monad
$\RMonad$ that the monoidal structure of the merge has been
realized and its basic properties analyzed.

Beside the theoretical motivations, monads have became a powerful
tool to implement non functional aspects into functional programming
languages, thanks to Moggi's original idea and to the work by
Wadler (see e.g. \cite{WadlerPhil1994a} and a series of papers
thereafter) and many others. It is now a day a common practice to
model imperative features into functional languages by means of
monads, especially by the community of Haskel programmers. This
relation to the programming practice is not by chance: 
among the basic motivations of the research
field we are about here is the desire
of methods for using efficiently classical logic principles to
develop programs whose adequacy to the specification can be
formally certified.

We observe in the main part of the paper that $\RMonad X$ is
isomorphic to $\State \arrow(X\times \State)$, that is isomorphic
as a type (though not as a monad) to the side-effect monad. But the most striking connections with the theme of
our work, is of course Moggi's monad of continuations, which after
Griffin's intuition, is used to type control
operators by G\"odel-Gentzen doubly negated types. The fact that
we do not find the continuation monad at the basis of our
construction is easily explained by the limitation to
1-backtracking we have put forward: we think that the use of the
full strength of continuations would give the possibility of
interpreting unbounded backtracking, but at the price of
loosing any intuition about the relation among classical proofs
and the interactive algorithms we could derive from them.

\section{Conclusions}
\label{IR_conclusions}

We have interpreted non-constructive proofs of arithmetical statements which can be obtained by using excluded middle over $\Sigma^0_1$ formulas as procedures that learn about their truth by redefining the value of Skolem functions. This process is at the same time an instance of two interpretations of classical logic: learning in the limit and  1-backtracking. The structure of proofs is reflected by their realizers, which are compositional, and parametric in the composition operation we call ``merge''. Realizers inhabit a computational type, hence a particular monad; actually monads are the structuring principle on which our construction relies.

As further steps of the presented research, we envisage the recasting of the (existing) extension of interactive realizers to $\HA+\EM$ in the framework of monads and, more importantly, the generalisation of interactive realizers to encompass {\bf EM$_n$} axiom schemata.

\end{document}